\documentclass[structabstract]{aa}

\usepackage{txfonts}
\usepackage{graphicx}
\usepackage{ulem}
\usepackage{natbib}

\begin{document}

\title{Giant Pulses with Nanosecond Time Resolution detected from the Crab Pulsar at 8.5 and 15.1~GHz}

\author{Axel Jessner\inst{1}
\and Mikhail V. Popov\inst{2}\thanks{e-mail: \tt{mpopov@asc.rssi.ru}}
\and Vladislav I. Kondratiev\inst{3,4,2}
\and Yuri Y. Kovalev\inst{2,1}
\and Dave Graham\inst{1}
\and Anton Zensus\inst{1}
\and Vladimir A. Soglasnov\inst{2}
\and Anna V. Bilous\inst{5}
\and Olga A. Moshkina\inst{2}}

\institute{Max-Plank Institute fur Radioastronomie, Auf dem Hugel 69, Bonn, Germany
\and Astro Space Center of Lebedev Physical Institute, Profsoyuznaya 84/32, Moscow 117997, Russia
\and Department of Physics, West Virginia University, 210 Hodges Hall, Morgantown, WV 26506, USA
\and Netherlands Institute for Radio Astronomy (ASTRON), Postbus 2, 7990 AA Dwingeloo, The Netherlands
\and Department of Astronomy, University of Virginia, P.O. Box 3818, Charlottesville, VA 22903, USA}
 

\abstract 
{
We present a study of shape, spectra and polarization properties of giant pulses (GPs)
from the Crab pulsar at the very high frequencies of 8.5 and 15.1~GHz. Studies at
15.1~GHz were performed for the first time.}
{
These studies were conducted to probe GP emission at
high frequencies and examine their intrinsic spectral and polarization properties 
with high time and spectral resolution. The high radio
frequencies also alleviate the effects of pulse broadening due to 
interstellar scattering which masks the intrinsic properties of GPs at
low frequencies. }
{
Observations were conducted with the 100-m radio telescope in Effelsberg in 
Oct-Nov 2007 at the frequencies of
8.5 and 15.1 GHz as part of an extensive campaign of multi-station multi-frequency 
observations of the Crab pulsar. A selection of the strongest pulses was recorded with 
a new data acquisition system, based on a fast digital oscilloscope, 
providing nanosecond time resolution in two polarizations in a bandwidth 
of about 500 MHz. In total, 29 and 85 GPs at longitudes of the main pulse and 
interpulse were recorded at 8.5 and 15.1 GHz during 10
and 17 hours of observing time respectively. We analyzed the pulse shapes, polarisation and 
dynamic spectra of GPs as well as 
the cross-correlations between their LHC and RHC signals.}
{
No events were detected outside main pulse and interpulse windows. 
GP properties were found to be very different for GPs emitted at longitudes 
of the main pulse and the interpulse. 
Cross-correlations of the LHC and RHC signals show regular patterns in the frequency domain for the 
main pulse, but these are missing for the interpulse GPs.
We consider consequences of application
of the rotating vector model to explain the apparent smooth variation 
in the position angle of linear polarization for main pulse GPs. 
 We also introduce a new scenario of GP generation as a direct consequence 
of the polar cap discharge.}
{We find further evidence for strong nano-shot discharges in the magnetosphere of the Crab pulsar. 
The repetitive frequency spectrum seen in GPs at the main pulse phase 
is interpreted as a diffraction pattern of regular structures in the emission region. The interpulse
GPs however have a spectrum that resembles that of amplitude modulated noise. Propagation effects
may be the cause of the differences.
}


\titlerunning{Giant Pulses with Nanosecond Time Resolution at 8.5 and 15.1~GHz}
\authorrunning{Jessner et al.}

\maketitle

\section{Introduction}
\label{int}
Radio giant pulses (GPs) represent the most striking phenomena of pulsar
radio emission. GPs were detected from the Crab pulsar \citep{staelin1968}, 
from the millisecond pulsar B1937+21 \citep{backer1984}, and 
from several other pulsars \citep{romani2001, johnston2003, joshi2004, knight2005}. 
However, detailed properties of GPs have been investigated only for those 
from the Crab pulsar \citep{lundgren1995, cordes2004} and from 
 B1937+21 \citep{kinkhabwala2000, soglasnov2004}.
Giant pulses demonstrate very peculiar properties, such as
i) very high flux densities 
exceeding $10^6$~Jy \citep{soglasnov2007};
ii) ultra short durations (few microseconds) with occasional bursts
shorter than 0.4~ns \citep{hankins2007};
iii) a power-law intensity distribution in contrast to the Gaussian distribution
for normal single pulses \citep{argyle1972, lundgren1995, popov_stappers2007};
iv) a very high degree of polarization \citep{cognard1996, popov2004, hankins2003};
v) a narrow range of longitudes and a particular relation with the
position of the components of the average profile \citep{soglasnov2004}.
These properties of GPs indicate that they represent the fundamental
elements of the emission and may provide direct insights into
the physics of the radio emission process. Indeed, other properties of pulsar emission,
such as the  radio spectrum can be easily modeled as being the result
of a very short timescale emission process \citep{Loehmer2008}.

GPs from the Crab pulsar have been detected in a broad frequency range
from 23~MHz \citep{popov2006b} to 8.5~GHz \citep{moffett1996, cordes2004, jessner2005};
No previous observations at higher frequencies have been reported except for the 
mention of a detection of a single GP at 15~GHz in a test observation by
\citet{hankins2000}.
The radio emission of the Crab pulsar at high radio frequencies is quite remarkable.
Its average profile 
shows a peculiar evolution with frequency \citep{moffett1996, cordes2004}.
According to these authors, at frequencies higher than 2.7~GHz the average profile consists of up to six
components at different longitudes, in particular, two additional high-frequency components (HFC1/HFC2)
appear between main pulse (MP) and interpulse (IP).  They are broad and have completely different properties
to the MP and IP.
\citet{jessner2005} reported the detection of GPs at the longitudes of HFC1 and HFC2 components,
while \citet{cordes2004} did not detected any GPs in these components
in their observations at the adjacent frequency of 8.8~GHz despite good sensitivity.
The main pulse, the strongest component of the Crab pulsar average profile at low frequencies, is 
hardly visible at high frequencies. The interpulse first vanishes completely at frequencies $>2.7$~GHz 
but re-appears above 4~GHz.
Its longitude, however, is shifted by $10\degr$ ahead
of that of the low-frequency IP.
\citet{hankins2007} believe that this high-frequency IP is a new component, 
distinct from the ``old'' low-frequency IP.
However, \citet{popov2008} find a close relation 
between low- and high-frequency IPs.

To approach an understanding of the nature of the GP radio emission mechanism,
it is important to study the intrinsic properties of individual GPs in detail and measure 
their true width, amplitude and polarization.
At low radio frequencies propagation
effects, in  particular interstellar scattering, obliterate the intrinsic pulse structure. 
As the scattering timescale decreases 
$\propto \nu^{-3.5}$ for the Crab pulsar only higher frequencies \citep{Kuzmin2008}
allow the observation of the very short time structures present in GPs. 
In this paper we present results of the analysis of GP properties measured at 8.5 and
15.1~GHz with the Effelsberg 100-m radio telescope. 
The details of the radio observations are described in Sect.~\ref{sec:obs}.
In Sect.~\ref{sec:red} we provide details of our data processing. Results, including GP
statistics, instantaneous radio spectra and polarization of GPs are presented in Sect.~\ref{sec:res}.
Finally, in Sects.~\ref{disc} and \ref{summary} our results are discussed and summarized.

\section{Observations}\label{sec:obs}
The observations were carried out at the two observing frequencies of 8.5 and 15.1~GHz 
 using the 100-m radio telescope in Effelsberg in a coordinated observing session with several other observatories.
The epoch and other details of the observations are summarized in Table~\ref{tab:obspar}.

\subsection{Receiver characteristics}
The 8.5 GHz HEMT receiver 
has a typical zenith system temperature of 27~K  
and a sensitivity of 1.35~K/Jy. The 15.1 GHz system is also a HEMT receiver, 
having a typical system temperature of 50~K and a sensitivity of 
1.14~K/Jy. The receivers were tuned 
to sky frequencies 8.576 and 15.076~GHz. 
Both receivers have circularly polarized feeds. Broadband signals with a bandwidth 
of 1~GHz for the 8.5~GHz system and 2~GHz for the 15.1~GHz system were detected at the receiver 
and used for the continuum calibration procedure. These signals were also fed to the Effelsberg 
pulsar observation system (EPOS) and used to monitor the signal quality. 
A remote controlled switched noise diode is built into each receiver for calibration purposes.

\subsection{Data acquisition}
The receivers used for the observations provide an IF signal (VLBI--IF) ranging 
from  0.5 to 1~GHz (effective bandwidth of 500~MHz).
This signal was fed to the Effelsberg Mark5 VLBI recording system for the simultaneous multi-frequency 
Mark5 experiment described in \citet{Kondratiev2010}.
The results presented here were obtained  using a Tektronix DPO~7254A digital storage
oscilloscope which recorded single strong GPs. The principles of the technique have been described 
in \citet{hankins2003}. 
The VLBI--IF signals corresponding to right-hand circular (RHC) and left-hand circular (LHC) 
polarizations were directly connected to 
the inputs (channel 1 and channel 2) of the oscilloscope. 
The 
12.5~Msamples (8-bit)  were recorded for each channel at a rate of 
2.5~Gsamples s$^{-1}$ giving a time window of 5~ms around the trigger epoch. 
A trigger signal was derived from the RHC IF signal by detection in a square-law detector.
The detector output was low-pass filtered using a commercial HP5489A filter unit which 
had a cut-off at 10~kHz. That signal was supplied to channel 3 of the DPO~7254A 
which was set to  trigger on the falling flank at a level exceeding by  5--6 times
the typical fluctuation in the signal. 
At this threshold setting one third of records were real GPs, and the rest were the result of 
 false triggers caused by noise.    
The time constant of the low-pass filter roughly corresponds to dispersion
smearing (320 and $70~\mu$s at 8.5 and 15.1~GHz respectively). 
Note that because of the dispersion broadening, the triggering is more sensitive
to the total power than to the peak amplitude of pulses.
Strong but very narrow pulses (e.g. single unresolved spike, ``nanoshot'')
are unlikely to be detected, while broad pulses of moderate peak strengths were captured.
Triggering the recording from only one low pass filtered circular 
polarization channel will however not cause a loss of any detections as GP occurring in only one 
circular polarization are practically unknown.

The oscilloscope operated in single-shot mode. For the first three 
sessions we had to save the data manually on the internal disk drive and re-arm the trigger.
In the last session at the observing frequency of 8.5~GHz we used an additional software 
module to automatically trigger and record the signals. There was, however, 
no significant difference in the amount of 
data recorded between the sessions, so that the manual operation did not entail any 
significant loss of data.

\begin{table*}
\caption{Observing setup and processing details.
The table lists the observing frequency, $f$; the date of observations; the bandwidth, $B$; 
system equivalent flux density (SEFD); observing time, $T_\mathrm{obs}$;
instrumental delay between two polarizations channels, $\delta t$; offset correction
of the observed position angle of linear polarization, $\Delta\phi_\mathrm{PA}$;
number of GPs found, $N_\mathrm{GP}$; peak flux density of the strongest GP, $S_\mathrm{max}$.} 
\label{tab:obspar}
\centering
\begin{tabular}{c c c c c c c c c c}
\noalign{\smallskip}\hline\hline\noalign{\smallskip}
$f$ & Date & $B$ & SEFD & $T_\mathrm{obs}$ & $\delta t$ & $\Delta\phi_\mathrm{PA}$ & $N_\mathrm{GP}$ & $S_\mathrm{max}$ \\
(GHz)    & (2007)& (MHz)& (Jy)& (h)& (ns)& ($\degr$)& & (kJy)\\
\noalign{\smallskip}\hline\noalign{\smallskip}
15.1 & Oct 24--25 & 500 & 150 & 6& 2.5 & 78& 42 & 60\\
$\ldots$ & Nov $\!\phantom{2}$4--5$\phantom{0}$ & 500 & 150 & 11& 2.5 & 78 & 43 & 40\\
8.5  & Nov $\!\phantom{2}$9--10 & 400 & 110 & 10& 5.5 & 143 & 29 & 150\\
\hline
\end{tabular}
\end{table*}

\section{Data reduction}\label{sec:red}

Preliminary inspection of the recorded data entailed Fourier transforming the whole 
data array, applying coherent dedispersion 
and creating the signal power envelope as well as the dynamic spectrum. The dynamic spectrum enabled 
us to discriminate easily between dispersed
GPs from the source and undispersed interference pulses or noise events. 

\subsection{Flux density calibration}

The signal of the noise diode was first calibrated as the result of 
cross-scans on continuum sources 3C48, 3C286, and DA193.  
According to the catalog of galactic supernova remnants \citep{green2006} 
the Crab nebula has a flux density of 1040~Jy at 1~GHz coming 
from an area of 35 square arc minutes with a spectral index of  0.3 at the center.
Combining the spectral decay and the beam widths of $80\arcsec$ at 8.5~GHz, and of $50\arcsec$ at 15.1~GHz,
it resulted in nebula contributions of 114 and 40~K, respectively.  
\citet{Baars1977} noted earlier
 that the nebula is a good calibrator for single-dish radio telescopes
because of its high signal. Thus, sky, nebula and receiver contributions resulted in background temperatures of 140-150~K at 8.5~GHz and 90-100~K at 15.5~GHz, or corresponding
system equivalent flux densities (SEFD) of 110 and 150~Jy (see Table~\ref{tab:obspar}).

In an extra calibration step before the start the calibration signal was switched  continuously 
twice per second with 1-ms duration and recorded
by the Tektronix oscilloscope to provide a reference for the signal strength in the recording equipment. 
The results for the Tektronix detection system were 
found to be consistent with the above estimates, giving  
an effective trigger threshold of 600 to 1000~Jy with a time constant 
of $100~\mu$s for the detection system. As typical nanopulse clusters have a duration 
of a few $\mu$s, we effectively detected GPs only when
their peak emission exceeded about 10~kJy. 

\subsection{Coherent dedispersion and envelope detection}\label{sub:dedisp}

With the sampling rate of 2.5~GHz used in the Tektronix DPO~7254A
one obtains an IF spectrum ranging from 0 to 1.25~GHz. 
However, the recorded signal occupies only the range of 500--900~MHz at 8.5~GHz, 
and 500--1000~MHz at 15.1~GHz, thus providing effective time resolution of about 2.5 and 2~ns,
respectively.
The dispersion smearing time ($\tau_\mathrm{DM}$) for the Crab pulsar 
at 8.5~GHz  and 400~MHz bandwidth is about 320~$\mu$s, 
or 800\,000 samples at 0.4~ns sampling rate. 
For the  coherent dedispersion one requires a minimum time interval $T$ greater 
than $2\tau_\mathrm{DM}$,
and we used number of samples $N = 2^{21}$ corresponding
to a time interval $T = 838.86~\mu$s.
To recover the signal over the whole Tektronix record ($N=12\,500\,000, T=5$~ms) 
we successively used short portions with $N=2^{21}$ overlapping by $2^{20}$ samples. 
Only the last half of each portion was used, the first half was padded with zero 
as the first 800\,000 points 
are spoiled because of the cyclic nature of the Fast Fourier
Transform (FFT) convolutions. 
We applied the same procedure in each frequency range in spite 
of the fact that $\tau_\mathrm{DM}$  is much smaller at 15.1~GHz.

Intrinsic receiver parameters and the frequency-dependent cable
attenuation between receiver cabin and recording room  result in a
bandpass that  is far from flat in any frequency range.
A bandpass correction was implemented simultaneously
with the dedispersion routine. Templates for the correction 
specific to each frequency ranges and polarization channel
 were obtained by smoothing the average power spectra calculated over the 
off-pulse portions of the original records.
To form the edges of the corrected bands we used a simple 
attenuation function $Y_i=0.5[\cos(\phi_i-\pi)+1]$ with 
$\phi_i$ changing from zero at the beginning to $\pi$ 
at the end of the corrected portion of the passband occupying 
about $1.5\%$ (5000 harmonics) of the whole bandwidth.

To bring the sampling time into a reasonable correspondence 
with the bandwidth of the recorded signal we restricted the Inverse Fourier Transform (IFT) 
to only half of the spectrum 
(400\,000 harmonics), leading to a new sample interval of 0.8~ns. 

To calculate the observed Stokes parameters (see Sect.~\ref{sub:pol}) 
it is necessary to remove any differential delay $\delta t$ between the voltages recorded
in the two polarization channels caused by unavoidable differences in
feed characteristics and signal cabling. The values of the 
delay were found from average cross-correlation functions (CCF) between
signals recorded in RHC and LHC channels on the off-pulse portions of the Tektronix files. 
Because of the considerable linear polarization of the radio emission from the 
area of the Crab nebula in the vicinity of the pulsar, the CCFs show good maxima, 
the position of which compared to zero lag 
gives us the required value of the delay $\delta t$. 
The values of this delay for each observing session are presented in 
the Table~\ref{tab:obspar}, column~6.

To compensate for the delay $\delta t$ we applied a linear phase shift 
in the spectrum of one polarization channel  
with the slope $2\pi\delta t/T$. 
Thus, after the FFT of the recorded signal
of duration $T$
we multiplied each complex harmonic of the spectrum $Y_j$ by a 
complex multiplier $A_j\exp(i\phi)$, where $A_j$ is a bandpass correction function, 
and $\phi$ is given by the expression from \citet{hankins_rickett1975}:
$$
\phi=\frac{2\pi}{T}\left(j\delta t-\frac{j^2 \mathrm{DM}}{TDf_0^3[1+j/Tf_0]}\right)~,
$$
where DM is dispersion measure in pc cm$^{-3}$,
$D$ is dispersion constant equal to
$2.41\times10^{-16}$ 
pc cm$^{-3}$ s, 
$f_0$ is the frequency of the lower edge of the band, and $j$ is the 
current number of a given harmonic. 

For coherent instantaneous envelope detection we suppressed the negative frequencies before
calculating the IFT \citep{Bracewell}. This yields the analytical signal 
$u_a(t) = u(t) -i {\cal H}(u(t))$ 
corresponding to the dedispersed amplitudes $u(t)$, with ${\cal H}(u(t))$ as their
Hilbert transform. An analytical signal $u_a(t)=U(t)e^{i\omega t}$
can always be written as the product of a real-valued envelope function $U(t)$ and a complex
phase function. Simply squaring $u_a(t)$ then provides the 
instantaneous signal powers $P(t)=U^2(t)$. Coherent envelope detection was used throughout
our analysis as it provides the maximum time resolution that can truly be obtained from
our band-limited data. 

\subsection{Effective time resolution}
\label{sub:beam}
\begin{figure}[htbp]
\centering
\includegraphics[scale=0.4]{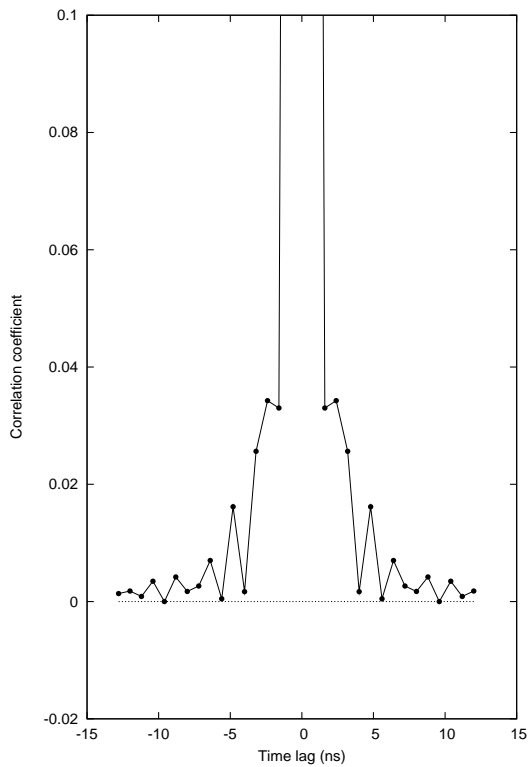}
\caption{Average off-pulse ACF at 15.1~GHz, illustrating the level of sidelobes.
To show the sidelobes in more details the ACF was cut off at the value of 0.1.
}
\label{fig:sidelobes}
\end{figure}

For a signal recorded in a frequency band $B$ the time resolution 
is equal to $1/B$. 
The exact form of the resolving profile 
depends on the shape of the receiver bandpass. Figure~\ref{fig:sidelobes} shows
the profile applicable to our observations at 15.1~GHz. It is
derived as the average autocorrelation function (ACF) calculated over off-pulse portions
of the reconstructed envelope. One can see that the first sidelobes of our resolving function
are at a sufficiently low level of about $4\%$.
Thus when attempting to restore the true pulse shape, any distortions or artefacts are, 
even for the strongest pulses, comparable to the noise level, and in most cases below noise. 

\subsection{Arrival Times}
\label{sub:toa}

The Tektronix recording system has an internal quartz clock that was aligned
with the station clock via internet. A timestamp derived from the internal clock 
when a trigger was received, was recorded with each data record.  
 The paucity of high resolution detections with the Tektronix system enabled us to identify
 individual GPs in the concurrent continuous Mark5 record and to determine the clock 
offsets and drift ($ \le 33\rm ns/s$) and as a consequence also the arrival times of all observed 
GPs. The GP trigger was derived from a broad--band total power detection followed by a low pass 
filter. That limited the timing accuracy of the trigger to about 1 ms when compared to Mark5. 
A higher precision  of the GP arrival times is possible by referring 
the offset for individual GP spikes to the trigger time. It was however 
not required for the purpose of our analysis.

\subsection{Polarization}
\label{sub:rpol}
To obtain the Stokes parameters $I,V,Q,U$ we used the following expressions:
$$I=W_rW_r+W_iW_i+X_rX_r+X_iX_i$$
$$V=W_rW_r+W_iW_i-X_rX_r-X_iX_i$$
$$Q=2(W_rX_r+W_iX_i)$$
$$U=2(X_rW_i-X_iW_r)$$
$$L=\sqrt{U^2+Q^2}, \phi_\mathrm{PA}=0.5\arctan(U/Q),$$
where $W_r, W_i, X_r, X_i$ are real and imaginary components for 
every sample of the analytical signal resulting from
coherent dedispersion in the RHC ($W$) and the LHC ($X$) channels.
As mentioned in Sect.~\ref{sub:dedisp}, the instrumental delay $\delta t$ between
RHC and LHC channels was removed during the dedispersion. 
To equalize the amplification in $W$-,$X$-channels we normalized the analytical
signal by the root-mean-square deviation (rms) of the off-pulse portions
of the records. These off-pulse parts of any record, including
records  produced as a result of false detection, were used to determine the
Stokes parameters for radio emission of the Crab nebula from the region
within the beam near the position of the pulsar B0531+21. 
Approximately one hundred off-pulse measurements were taken during every observing
session. The degree of linear polarization was found to be
approximately constant.
We corrected the observed position angle of linear polarization ($\phi_\mathrm{PA}$)
for changes in parallactic angle during the observations. 
Measured values of 
the degree of linear polarization were about 11\% and 7\% at 8.5 and 15.1~GHz,
respectively. Offset corrections for the observed positional angles $\Delta\phi_\mathrm{PA}$ 
were found to be
$143\degr$ and $78\degr$  for 8.5 and 15.1~GHz, correspondingly (see Table~\ref{tab:obspar}, column 7).
These last corrections are required to rotate the measured polarization to  $\phi_\mathrm{PA}=135\degr$ 
at both 8.5 and 15.1~GHz, the value taken from maps in \citet{wilson1972} for a position 
near the pulsar.
A reasonable agreement of our measurements with the published data on
the polarization properties of radio emission from the Crab nebula ensures 
that our calculated Stokes parameters reflect the true properties
of the detected GPs. The accuracy of our measurements of the Stokes parameters $I, Q, U$and $V$ is estimated to be about 10\%. 
{
Since $L=\sqrt{Q^2+U^2}$ is a positive definite 
quantity and therefore cannot have a zero mean, one has take into account an additional offset from
the  "noise contribution" of
$\Delta L = \sqrt{\pi \over 4}\sigma$, where $\sigma$ is the standard deviation 
of the off-pulse noise of the signals that comprise Q or U \citep{Davenport&Root}. 
It was found to be at a level of about 0.1\%  for smoothed by 80 ns single pulses 
considered in section 4.3. The noise offset is at a level of 0.7\% for nanoshots averaged over 4 ns, 
analysed in sections 4.3.1 and 4.3.2. In the Figs.~2-5, 
where examples of strong GPs are presented (unsmoothed signal), 
the bias constitutes less then 1\% of the GPs peak intensity. Therefore the noise correction 
is small in all cases
 and does not contribute significantly to the observed degree of linear polarization.   
}
\section{RESULTS}\label{sec:res}
\subsection{General statistics}\label{sub:stat}
GP arrival times were converted into pulse longitudes using 
the pulsar timing package TEMPO and the Crab pulsar ephemeris 
for the epoch supplied by Jodrell Bank \citep{JBE2008}. 
The millisecond accuracy of the Tektronix trigger times was sufficient to 
unambiguously identify the origin of the GP with the known features of 
the average pulse profile.

All pulses detected with the Tektronix DPO~7254A have longitudes 
corresponding to MP or IP. No pulses were identified as HFCs.
During about 10 hours of observations at 8.5~GHz we detected 29~GPs,  
of which only 5 originated at the longitude of the IP. 
In contrast, during two observing sessions at 15.1~GHz 
we detected 85~GPs, 
but only 7 occurred at the longitude of the MP. 
The peak flux density of 
the brightest GP at 8.5~GHz was about 150~kJy, and about 60~kJy for
the strongest components of GP structures seen at 15.1~GHz. 
The total number of detected
GPs is not large enough for a statistical study of their properties.
The complete statistical analysis using the continuous Mark5 data will be 
reported in \citet{Kondratiev2010}. In the following sections we 
will analyze the different properties of
individual GPs  originating at MP and IP longitudes.
\begin{figure*}[hbtp]
\centering
\includegraphics[angle=-90,scale=0.5]{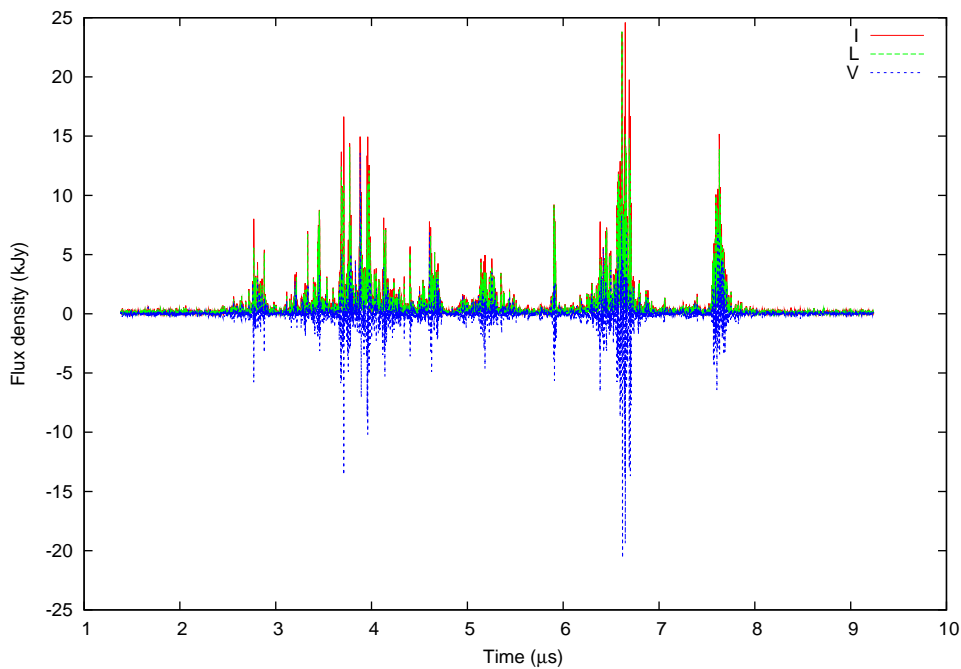}
\caption{Example of GP detected at 8.5~GHz shown with a sampling time of 0.8~ns. 
The pulse originates close to the MP longitude. The solid red line indicates the 
total intensity (I), the dashed green line~-- the linear polarization (L), and the dashed 
blue line~-- the circular polarization (V).} 
\label{fig:pulse557t}
\end{figure*}

\begin{figure*}[hbtp]
\centering
\includegraphics[angle=-90,scale=0.5]{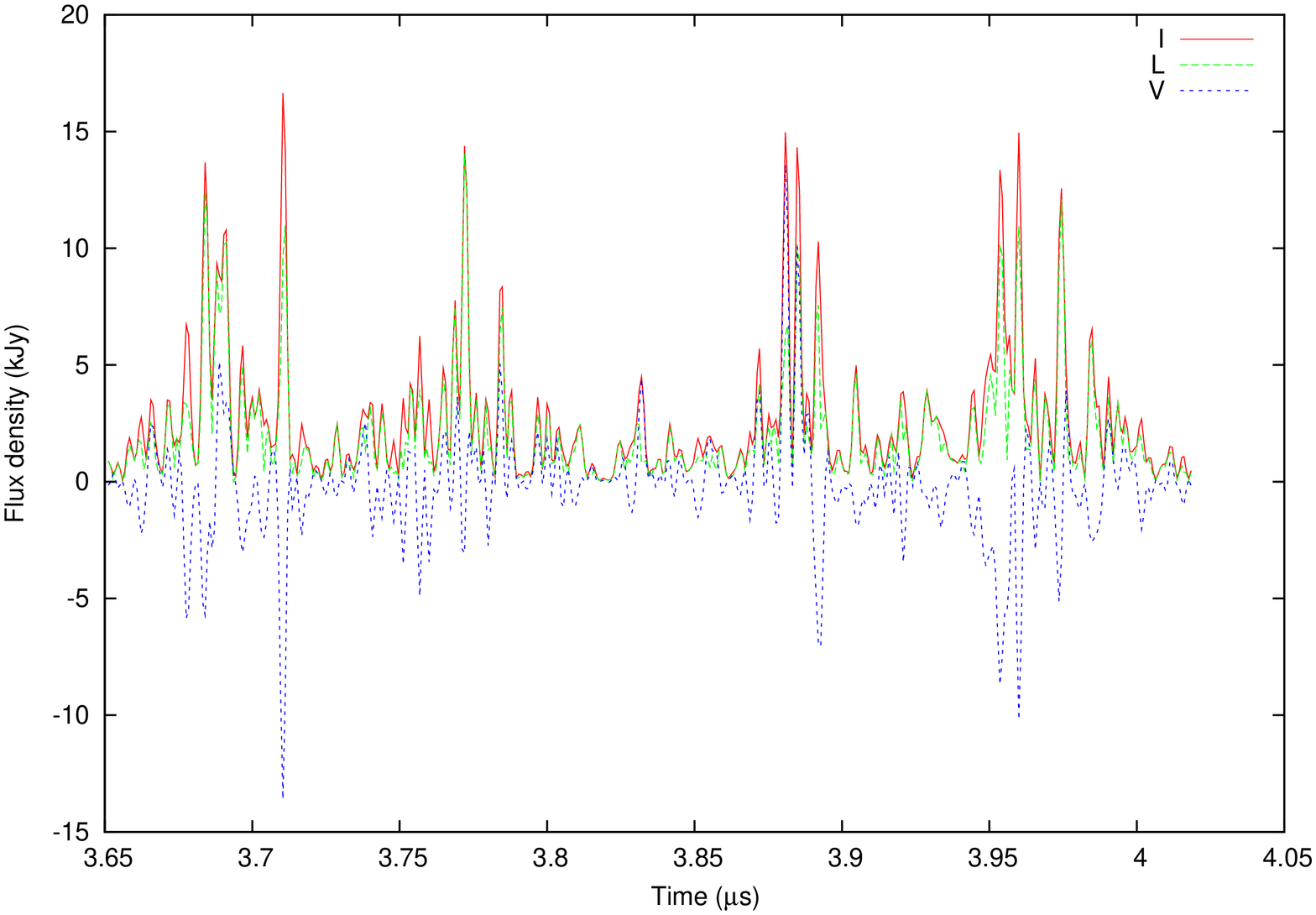}
\caption{Zoom-in portion of the GP shown in Figure~\ref{fig:pulse557t}. 
The solid red line indicates the total intensity (I), the dashed green 
line~-- the linear polarization (L), and the dashed blue line~-- 
the circular polarization (V).} 
\label{fig:pulse557p}
\end{figure*}

\begin{figure*}[hbtp]
\centering
\includegraphics[angle=-90,scale=0.5]{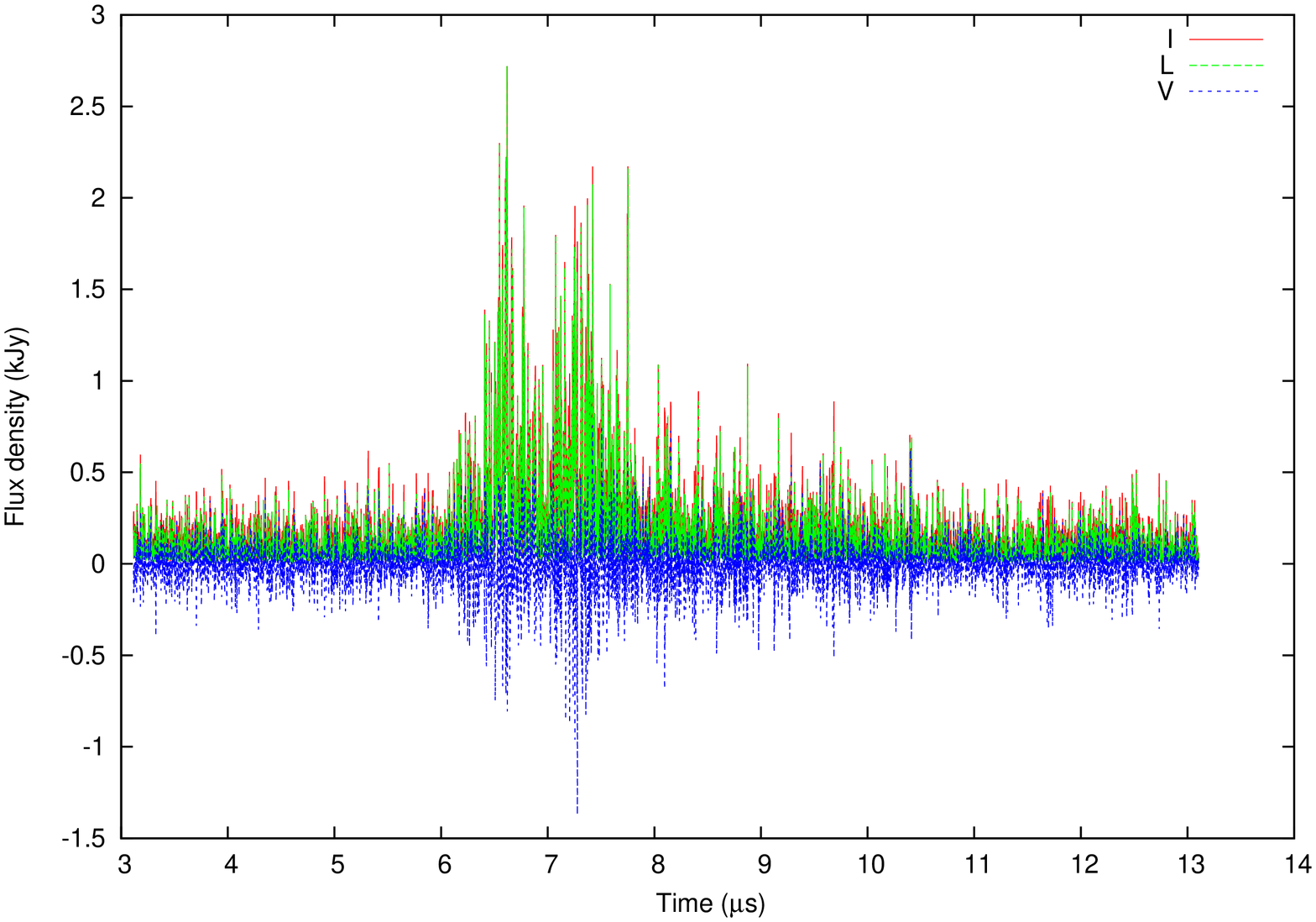}
\caption{Example of GP detected at 8.5~GHz shown with a sampling time of 0.8~ns. 
The pulse originates at IP longitudes. The solid red line indicates the 
total intensity (I), the dashed green line~-- the linear polarization (L), and the dashed 
blue line~-- the circular polarization (V).} 
\label{fig:pulse293t}
\end{figure*}

\begin{figure*}[hbtp]
\centering
\includegraphics[angle=-90,scale=0.5]{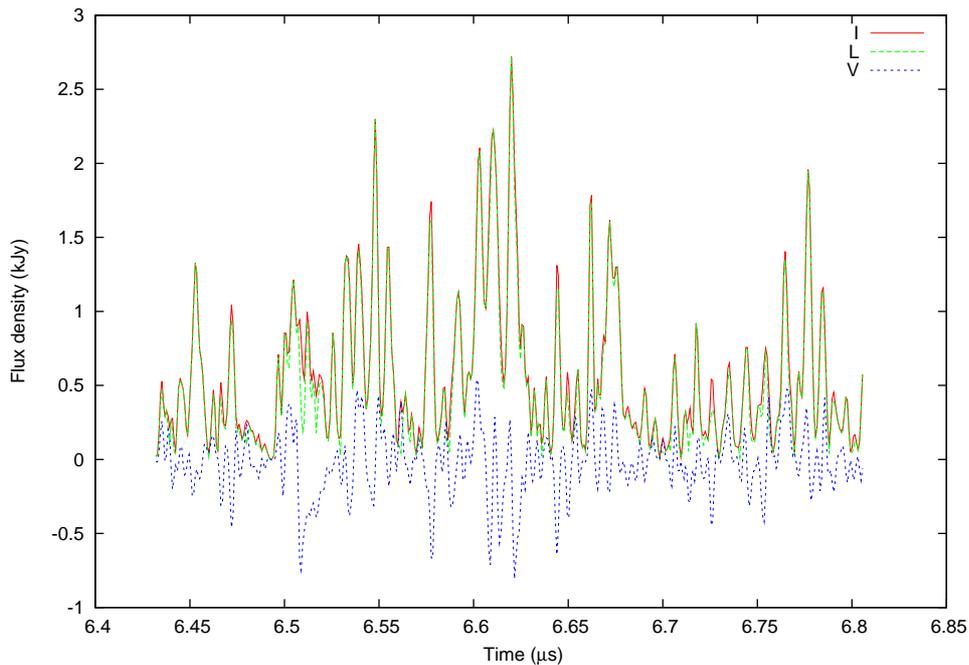}
\caption{Zoom-in portion of the GP shown in Figure~\ref{fig:pulse293t}.
The solid red line indicates the 
total intensity (I), the dashed green line~-- the linear polarization (L), and the dashed 
blue line~-- the circular polarization (V).}
\label{fig:pulse293p}
\end{figure*}

\subsection{Shapes of single pulses}\label{sub:shape}

Examples of strong GP profiles at 8.5~GHz both at the longitudes of MP and IP 
are presented in Figs.~\ref{fig:pulse557t}--\ref{fig:pulse293p}
simultaneously in total intensity ($I$), linear ($L$), and circular ($V$) polarizations.
One can see clear differences between shape and polarization of GPs
originating at longitudes of MP and IP. 
MP giant pulses (MPGPs, Figs.~\ref{fig:pulse557t},\ref{fig:pulse557p}) 
consist of several distinct groups of
``microbursts'' with strong unresolved ``nanoshots''\footnote{These two terms were 
originally introduced by \citet{hankins2007}} in every group, 
while IP giant pulses (IPGPs, Figs.~\ref{fig:pulse293t},\ref{fig:pulse293p}) 
demonstrate only a uniform increase of  noise over a certain time interval.
These differences were first reported by \citet{hankins2007}.
There is one exceptional MPGP with a smoother waveform which we discuss below 
in Sects.~\ref{sub:pol} and \ref{sub:pec}. The waveforms of all IPGPs are rather similar 
in shape, usually showing a short rise time of about $0.6\pm 0.2~\mu$s 
and gradual decay after maximum for about $2.5~\mu$s. 
In contrast, MPGPs are very different in shape.

\begin{figure*}[hbt]
\centering
\includegraphics[width=8cm,height=8cm]{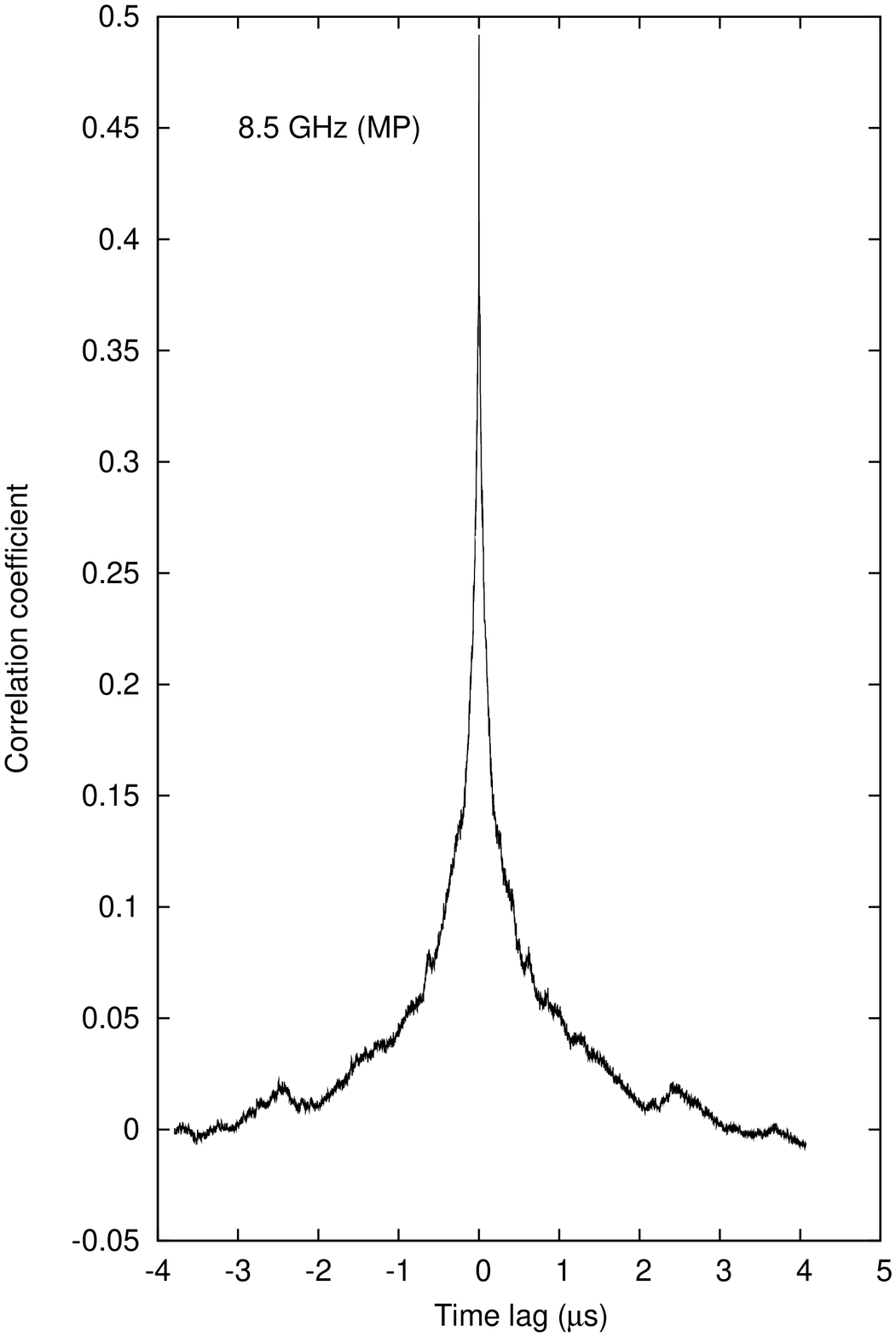}
\includegraphics[width=8cm,height=8cm]{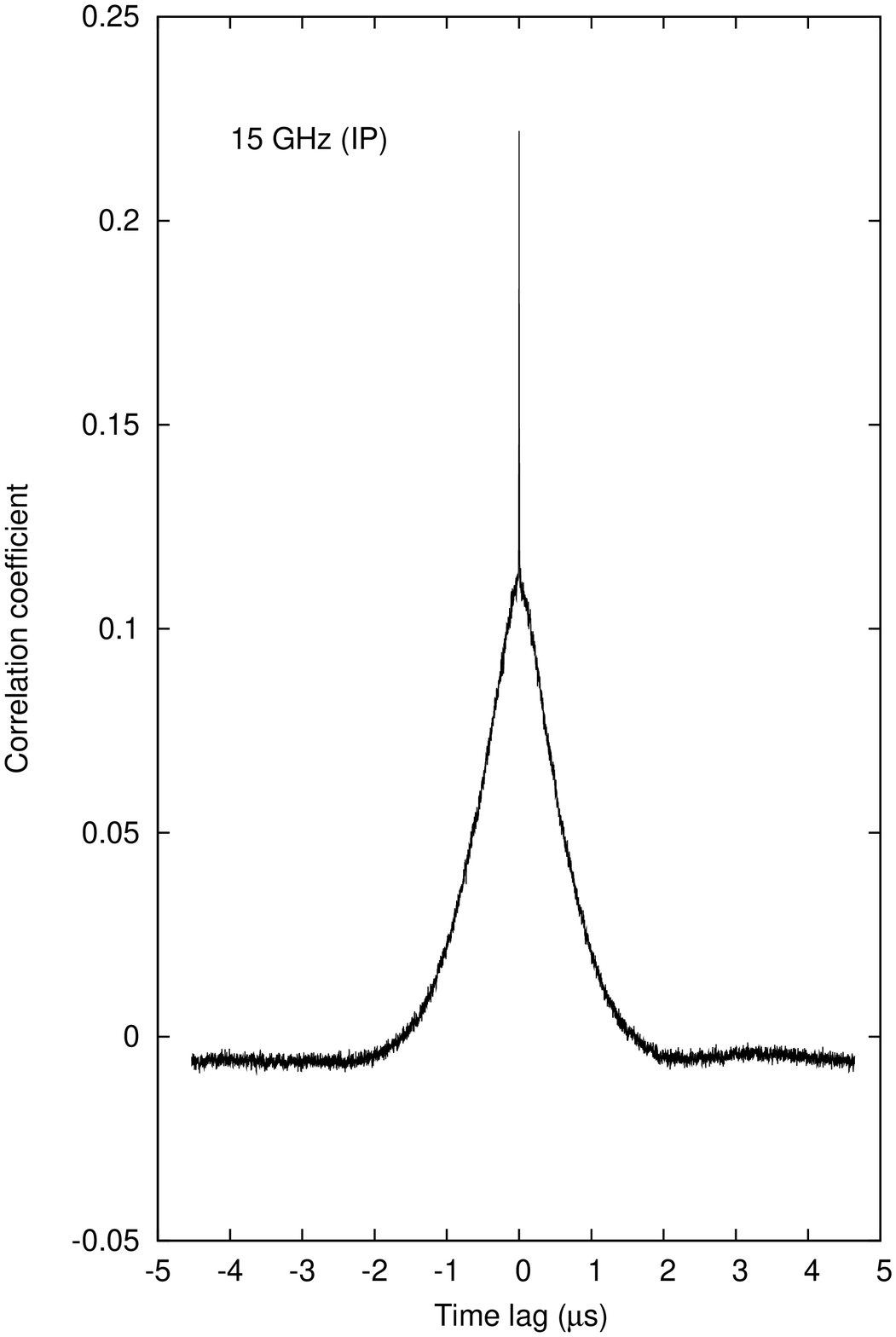}
\centering
\includegraphics[width=8cm,height=8cm]{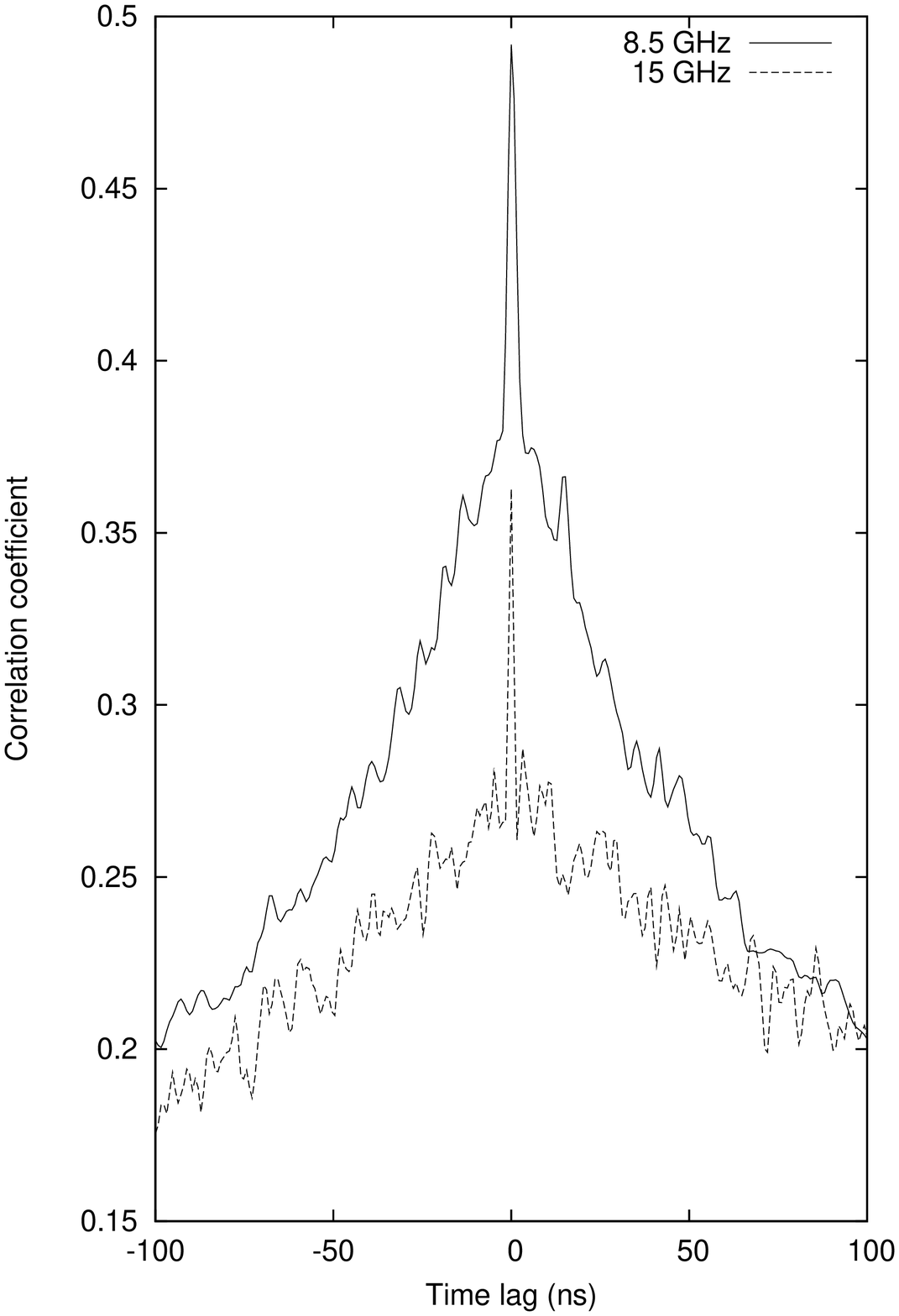}
\includegraphics[width=8cm,height=8cm]{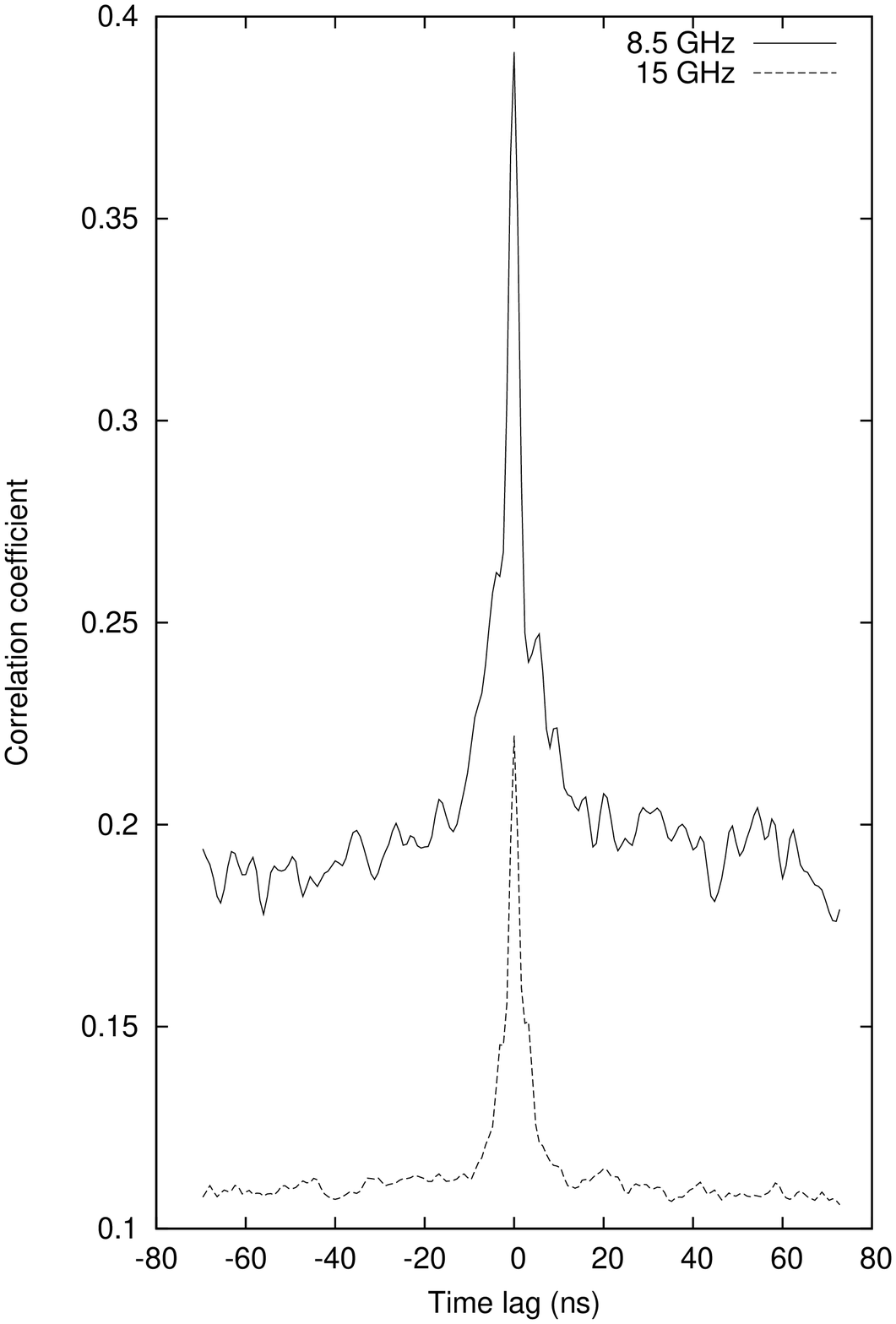}
\caption{Average CCFs between GP intensities detected in two polarization channels. 
The left panels represent MPGPs and the right panels correspond to IPGPs. 
Figures at the bottom show only the central part of CCFs.}
\label{fig:ccf}
\end{figure*}

MPGPs manifest a great variety of polarization
properties between different isolated microbursts exhibiting strong circular
polarization of both signs and strong linear polarization as well (see 
Fig.~\ref{fig:pulse557p}). On the other hand, all structural components of
IPGPs (as in Fig.~\ref{fig:pulse293p}) have a high degree of linear polarization.
The polarization properties of GPs will be discussed in more 
detail further in sect.~\ref{sub:pol}. 
Here we are concerned  only with the derivation 
of general parameters describing the observed pulse shapes. 
The traditional  method is the use of an ACF 
calculated for the on-pulse signal \citep{hankins1972,popov2002a}. 
However 
it is known that some GPs from the Crab pulsar have unresolved components (nanoshots)
with a duration shorter than 0.4~ns \citep{hankins2007}. 
The contribution of such components to an ACF would be undistinguishable 
from the impact of noise, both cause a spike at zero lag of the ACF.
A better alternative method is the use of the 
CCF between the on-pulse signals
received in two polarization channels with independent receiver noise.
In this case a narrow correlation spike at zero lag reflects the presence of linearly 
polarized unresolved components in the time structure of a signal received from a source.
The average CCFs are presented in Fig.~\ref{fig:ccf}.
The average CCF for MPGPs (Fig.~\ref{fig:ccf}, left) indicates the presence
of several time scales from a few nanoseconds to a few microseconds,
while the CCF for IPGPs (Fig.~\ref{fig:ccf}, right) shows only an unresolved spike at zero lag
and uniform Gaussian detail with a half-width of about $0.5~\mu$s.

Does the CCF spike at zero lag contains real unresolved nanoshots, or is it purely a result of noise?
The amplitude of isolated narrow pulses
 is very sensitive to the accuracy of the value of DM
used for coherent dedispersion \citep{rickett1975}. 
When processing with a range of assumed DM values, one may expect a rapid peaking of the zero-lag CCF spike 
near the correct DM in the presence of unresolved nanoshots,
and the absence of such a prominent maximum if only noise is present.

We adopt a value 
$a_0=D_\mathrm{max}-D_\mathrm{min}$ as amplitude of a zero-lag spike 
above the broader CCF shoulders, with $D_\mathrm{max}$ and $D_\mathrm{min}$ being 
maximum and minimum values in a series of $D_i$ calculated as 
$D_i=\mathrm{CCF}_{i+1}-\mathrm{CCF}_{i}$  in the range
of time lag $\pm\Delta t$ near the zero lag. 

\begin{figure}[hbt]
\centering
\includegraphics[width=8cm,height=10cm]{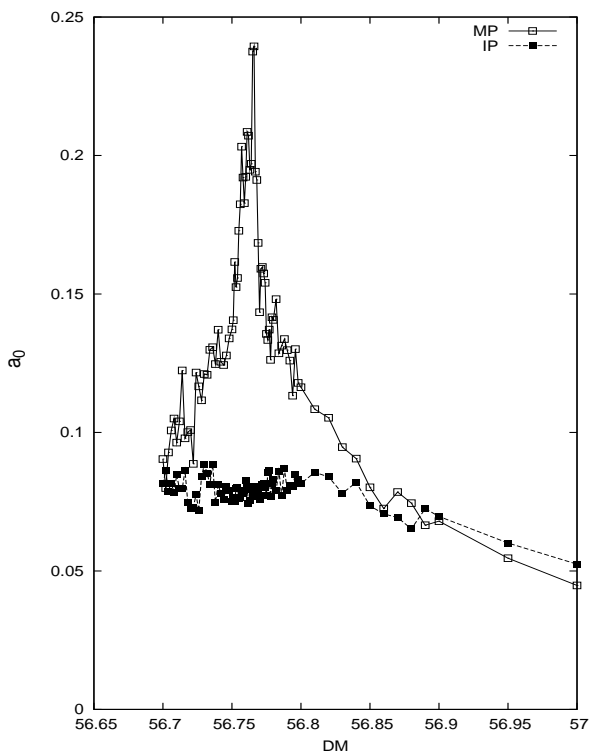}
\caption{
The dependence of the value of $a_0$ (see text) versus DM for MPGPs (open squares)
and IPGPs (filled squares).}
\label{fig:ccf_DM}
\end{figure}

The dependence of $a_0$ versus DM for MPGPs and IPGPs is shown 
in Fig.~\ref{fig:ccf_DM}.
Open squares represent the mean values of $a_0$ averaged
over 5 strong GPs detected at 8.5~GHz within the the MP longitude range.
Closed squares represent mean values of $a_0$ averaged
over 5 strong GPs detected at the same frequency at IP longitudes.
The curve for MPGPs has a very sharp maximum at $\mathrm{DM}=56.766$ pc cm$^{-3}$, 
which corresponds closely to the value published in the 
Jodrell Bank Crab Pulsar Monthly Ephemeris \citep{JBE2008} for our observing epoch.
The zero-lag peak of the CCF for IP data exists and has large amplitude, however
no marked dependence on assumed DM is found in this case.
Thus for the IP there is no evidence of large unresolved nanoshots which we
might identify with individual emitters.
It is interesting
to note that the central CCF peak for IP data has the same amplitude as
that of the extended CCF shoulders
in agreement with the amplitude modulated noise (AMN) model
proposed by \citet{rickett1975} to explain microstructure
in pulsar radio emission.

\subsection{Polarization of single pulses}
\label{sub:pol}
\begin{figure*}[hbt]
\includegraphics[width=5cm,height=6cm,angle=-90]{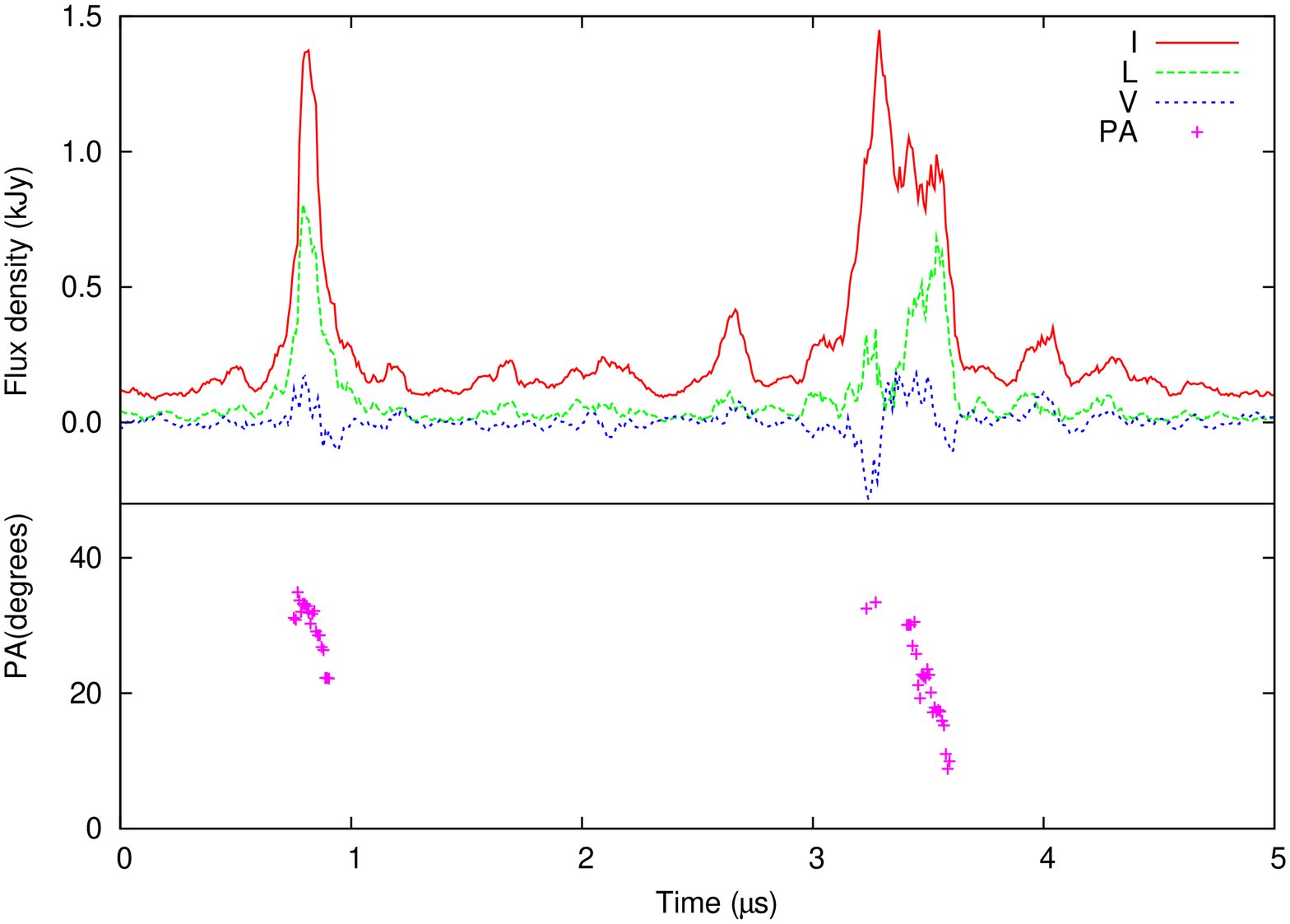}
\includegraphics[width=5cm,height=6cm,angle=-90]{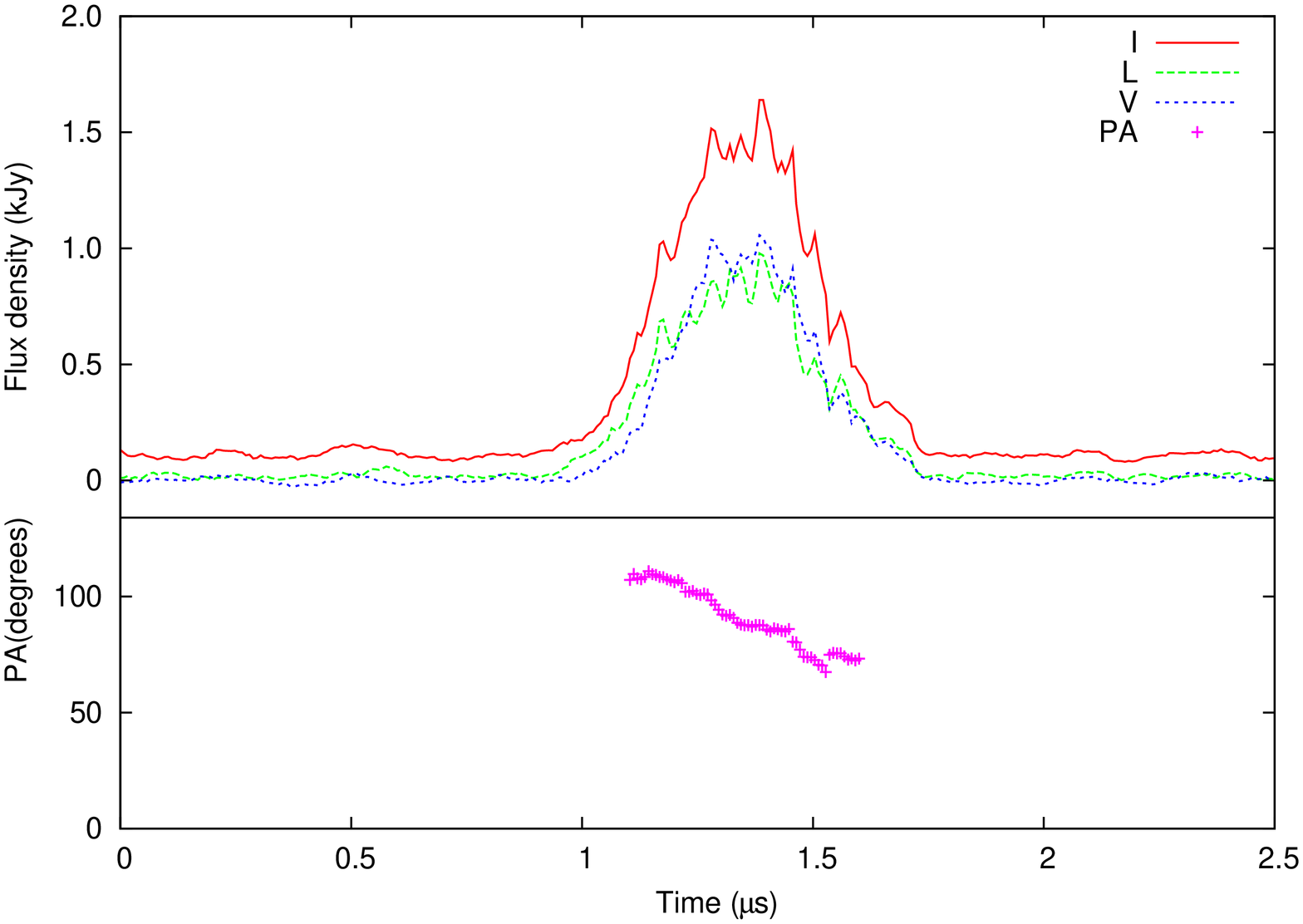}
\includegraphics[width=5cm,height=6cm,angle=-90]{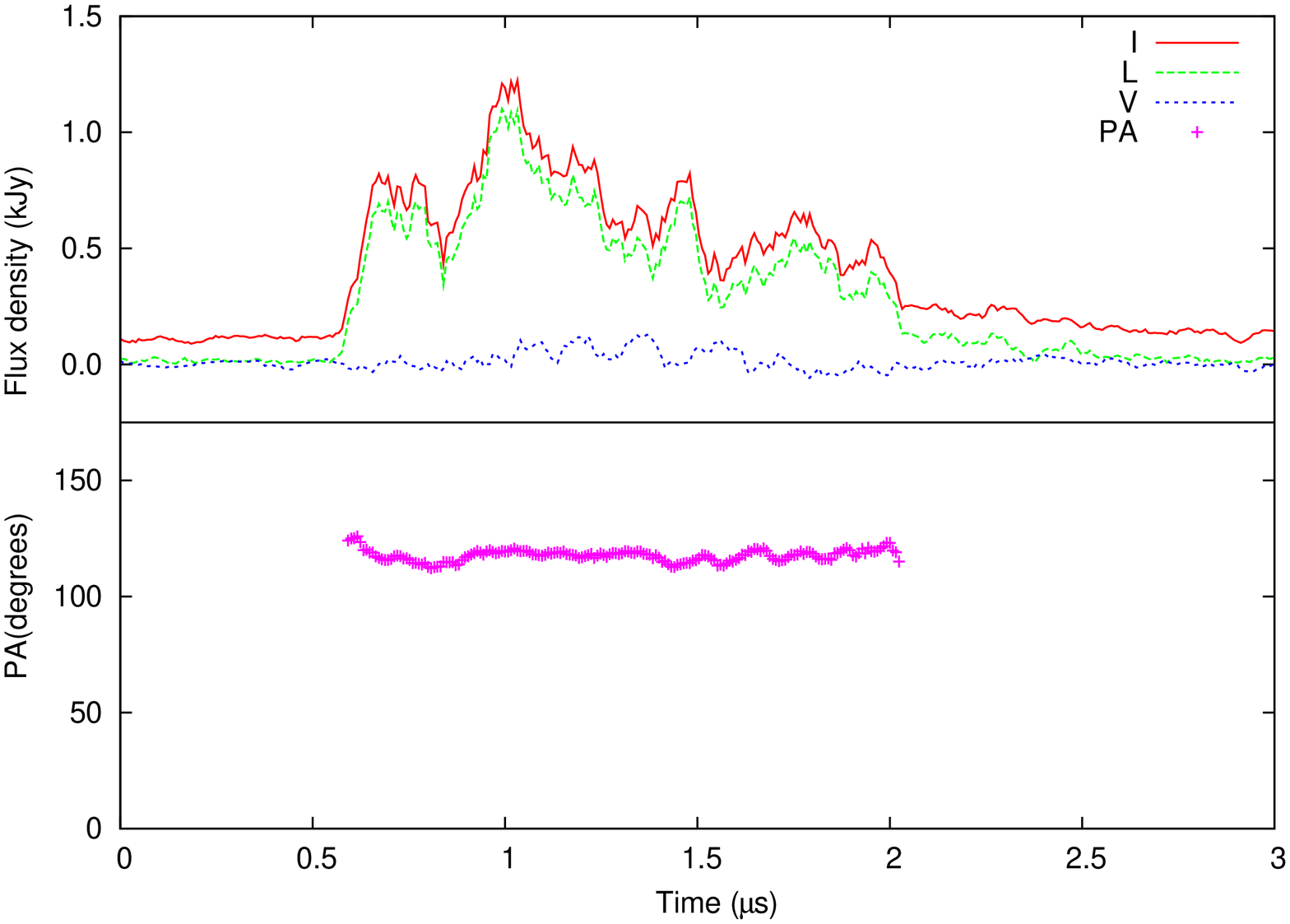}

\caption{
Smoothed profiles and position angle (PA) evolution for selected MPGP (left),
peculiar MPGP (middle, see Sect.~\ref{sub:pec}), and IPGP (right).
The IPGP shown was detected at 15.1~GHz, while the MPGPs were
found at 8.5~GHz. PA is in degrees (Y-axis), and intensity is in kJy.
The solid red line indicates the 
total intensity, the dashed green line~-- the linear polarization, the dashed 
blue line~-- the circular polarization, and black crosses~-- the PA variation.
}
\label{fig:smoothed_pol}
\end{figure*}

The calibration and processing described in section~\ref{sub:rpol} allowed us to present  
examples of GP profiles in Figs.~\ref{fig:pulse557t}--\ref{fig:pulse293p}
with full time resolution. They  show
a huge variety of polarization behavior, especially for GPs corresponding to
the MP longitude (see Fig.~\ref{fig:pulse557p}), 
where neighboring bursts may have opposite
signs of circular polarization. 
In an attempt to find trends in the observed parameters, 
particularly for the evolution of position angle (PA) of polarization, we first
applied running averaging with a bin size of 80~ns to 
all four Stokes parameters for each GP.
Examples of such smoothed GPs are shown in Fig.~\ref{fig:smoothed_pol}.
Left and right panels of the Fig.~\ref{fig:smoothed_pol} represent typical
profiles of MPGP and IPGP respectively.
The middle panel shows 
an unusual MPGP mentioned in Sect.~\ref{sub:shape}. 
Fig.~\ref{fig:smoothed_pol} shows the evident differences between 
polarization properties of MPGPs and IPGPs. 
An IPGP usually has a high degree of linear polarization (up to 100\% at the maximum)
and small variations of position angle, less than $\pm5\degr$ peak-to-peak. 

An MPGP shows a smaller degree of linear polarization (30--50\%),
and greater variations of PA ($\pm20$--$40\degr$). 
The PA evolves smoothly within separate components, with varying sweep rates and ranges.
These properties were confirmed for the majority of GPs. 

All IPGPs have similar properties, with one exception which we will discuss
briefly in Sect.~\ref{sub:pec}. The PA is largely constant, and it is about the same 
for all IPGPs at both frequencies, 8.5 and 15.1~GHz. 
In contrast to that, every MPGP is different. PA values for MPGPs
vary randomly from pulse to pulse, as well as within the single pulse,
in the whole range of 0--$180\degr$.

As discussed in Sect.~\ref{sub:shape}, 
MPGPs exhibit a great variety of shapes. 
They consist of several dense clusters (components) of microbursts
containing strong unresolved nanoshots (Fig.~\ref{fig:pulse557t} 
and left panel of Fig.~\ref{fig:smoothed_pol}). Within a 
distinct component or microburst, PA 
demonstrates 
a very regular smooth variation over a range of a few 
tens of degrees.
The PA sweep varies strongly from one component to another 
and may also have an opposite sign in neighboring microbursts. 

The MPGP shown in the middle panel of the Fig.~\ref{fig:smoothed_pol} has a very 
smooth, wide waveform without any clustering of microbursts, which is 
untypical for MPGPs. Perhaps it represents 
a different, 
uncommon category of MPGPs, therefore we discuss this pulse separately in
Sect.~\ref{sub:pec}. Regular PA variation is seen throughout the pulse 
duration rotating smoothly from $118\degr$ to $75\degr$.
Another remarkable feature of this particular GP is a surprising coincidence of its intensity 
in linear and circular polarizations.

Smooth regular variations of PA seen in MPGPs are
similar to those typical of
integrated profiles of many pulsars, 
which show a classical S-shape swing \citep[see, e.g.,][]{vonbaron1997}. It is tempting to apply 
the geometric rotating vector model \citep[RVM,][]{rvm}, commonly adopted in studying 
polar cap geometry, to explain PA evolution in MPGPs. This is
considered in Sect.~\ref{disc}.

\subsubsection{Polarization of nanoshots}
Further, we tried to obtain the average
polarization profile of isolated unresolved spikes (nanoshots) present
in GPs. 
All such spikes with intensity above a given threshold
were aligned in time by their maxima and added together. 
To avoid effects of variations of PA with GP duration, the intensity
in linear polarization was averaged using the magnitude of $L=\sqrt{Q^2+U^2}$.
Also, the signal was smoothed by 5 samples to reduce noise fluctuations. 

\begin{figure*}[ht]
\includegraphics[width=5cm,height=6cm,angle=-90]{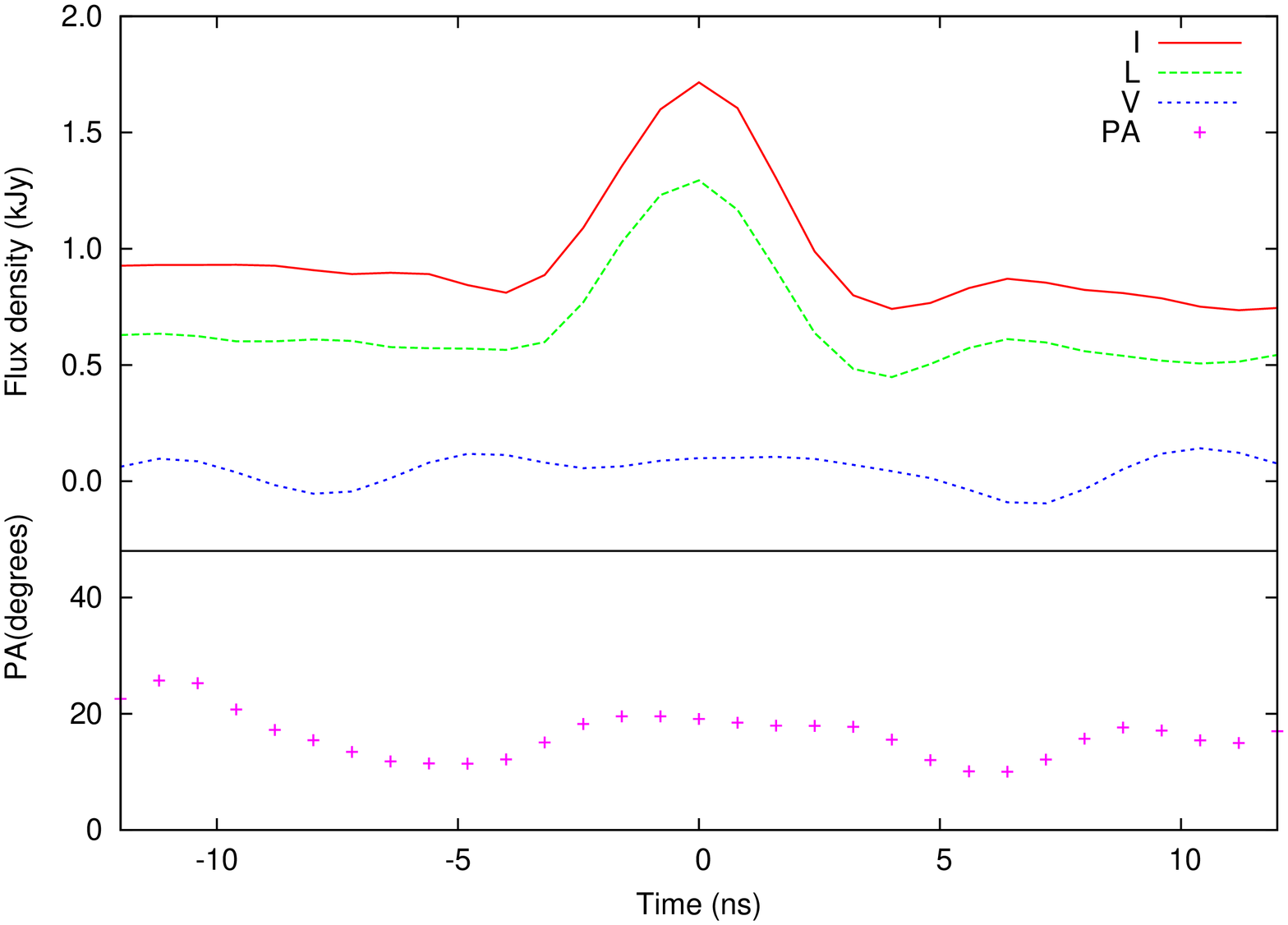}
\includegraphics[width=5cm,height=6cm,angle=-90]{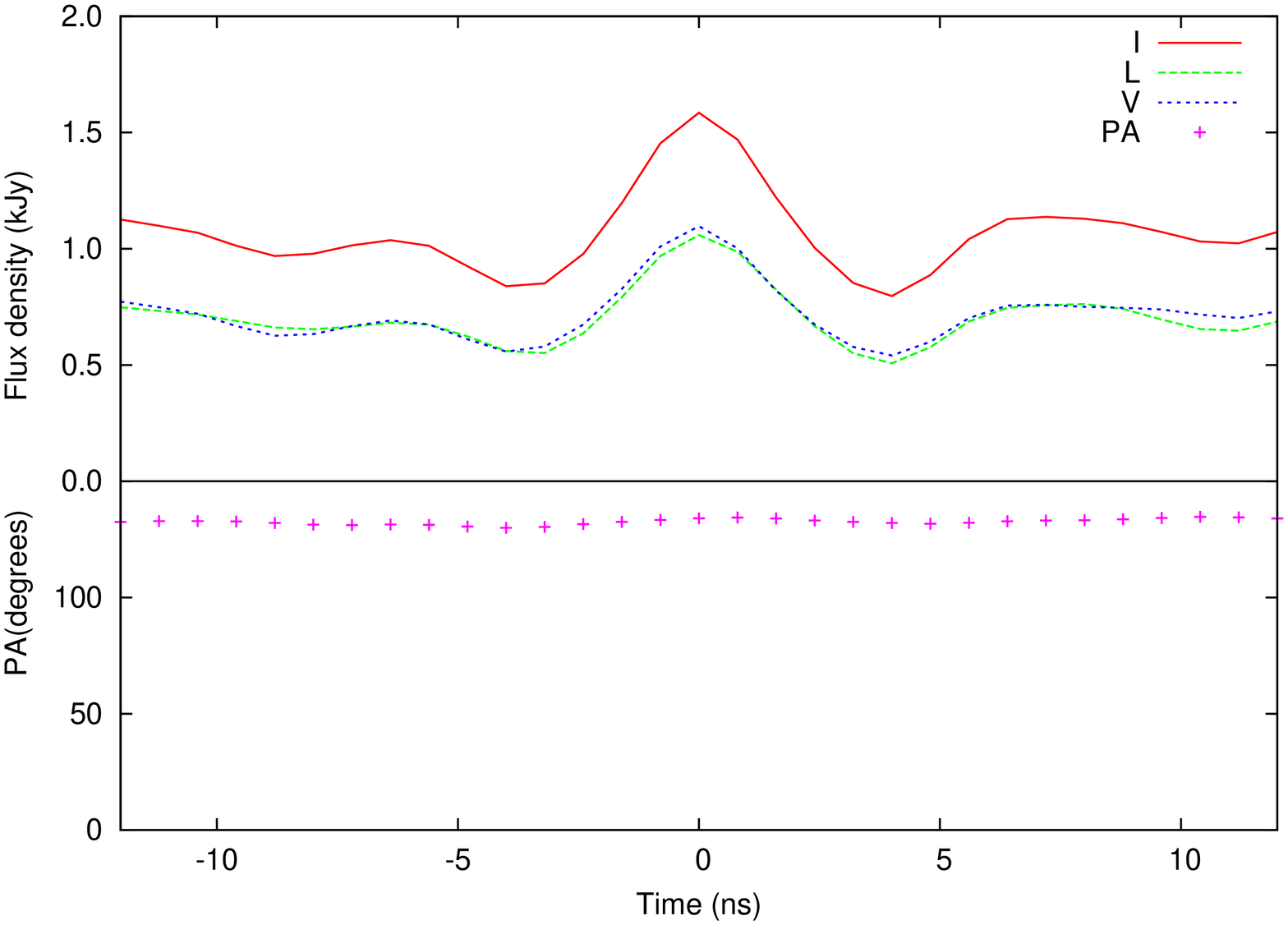}
\includegraphics[width=5cm,height=6cm,angle=-90]{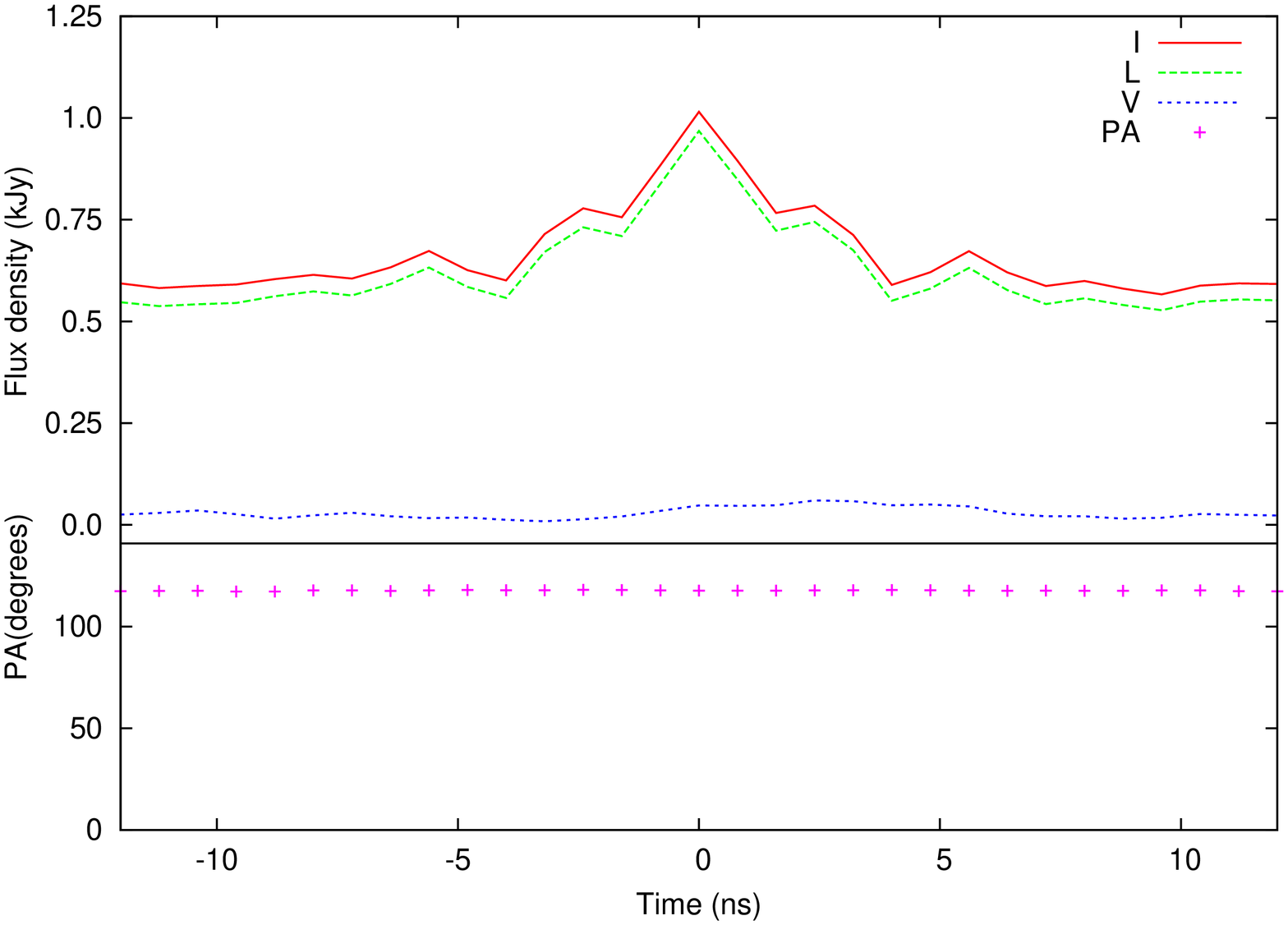}
\caption{Average profiles and position angle (PA) evolution of strong isolated spikes 
which are taken from the corresponding GP from Fig.~\ref{fig:smoothed_pol}.
PA is in degrees (Y-axis), and intensity is in kJy.
The solid red line indicates the 
total intensity, the dashed green line~-- the linear polarization, the dashed 
blue line~-- the circular polarization, and black crosses~-- the PA variation.}
\label{fig:spikes}
\end{figure*}

Typical GPs contained from  several
dozens up to hundreds of strong isolated spikes.
Examples of polarization profiles of such spikes are
presented in Fig.~\ref{fig:spikes}.

We hoped to find a regular behavior of PA or circular polarization ($V$)
through the average profile of a strong isolated emission spikes. However,
no general dependence for all GPs was found, not even for neighboring subsets 
of GPs. Nevertheless, Figs.~\ref{fig:smoothed_pol} and \ref{fig:spikes}
present examples of typical MPGPs (left and the middle panels),
and IPGPs (right panel).
The average profile for spikes belonging to MPGPs 
(Fig.~\ref{fig:spikes}, left) indicates that these nanoshots are truly isolated
(no other components are seen within at least ten nanoseconds). 
Such nanoshots have an approximately equal degree of circular and linear polarization.
However on average their Stokes parameter $V$ is close to zero because of the nearly equal
contributions from RHC and LHC components. 
The average position angle does not show any regular variation 
within the duration of a nanoshot.

The central panels in Figs.~\ref{fig:smoothed_pol} and \ref{fig:spikes}
represent polarization properties of the peculiar MPGP.
The profile of average spikes also reveals a remarkable feature of this GP: 
the degree of linear polarization is equal to the degree
of circular polarization, and the majority of spikes have circular polarization 
of the same sign (RHC). In fact, values of $L$ and $V$ stay astonishingly 
close together over the whole pulse duration.
They also evolve together in the average profile of outstanding spikes (see Fig.~\ref{fig:spikes}).
This average profile indicates the presence of close satellites near the majority
of spikes with similar polarization properties. The PA varies slightly.

The polarization properties displayed in the right panels
of Figs.~\ref{fig:smoothed_pol} and \ref{fig:spikes} correspond to the IPGPs.
A high degree of linear polarization is evident in both the smoothed pulse profile
and in the average profile of spikes. This profile contains of 
many small spikes merged together.
Also, the profile confirms a constancy of PA 
for IPGP.

\subsubsection{Polarization statistics of nanoshots}
\begin{figure*}[ht]
\centering
\includegraphics[width=6.5cm,height=6cm]{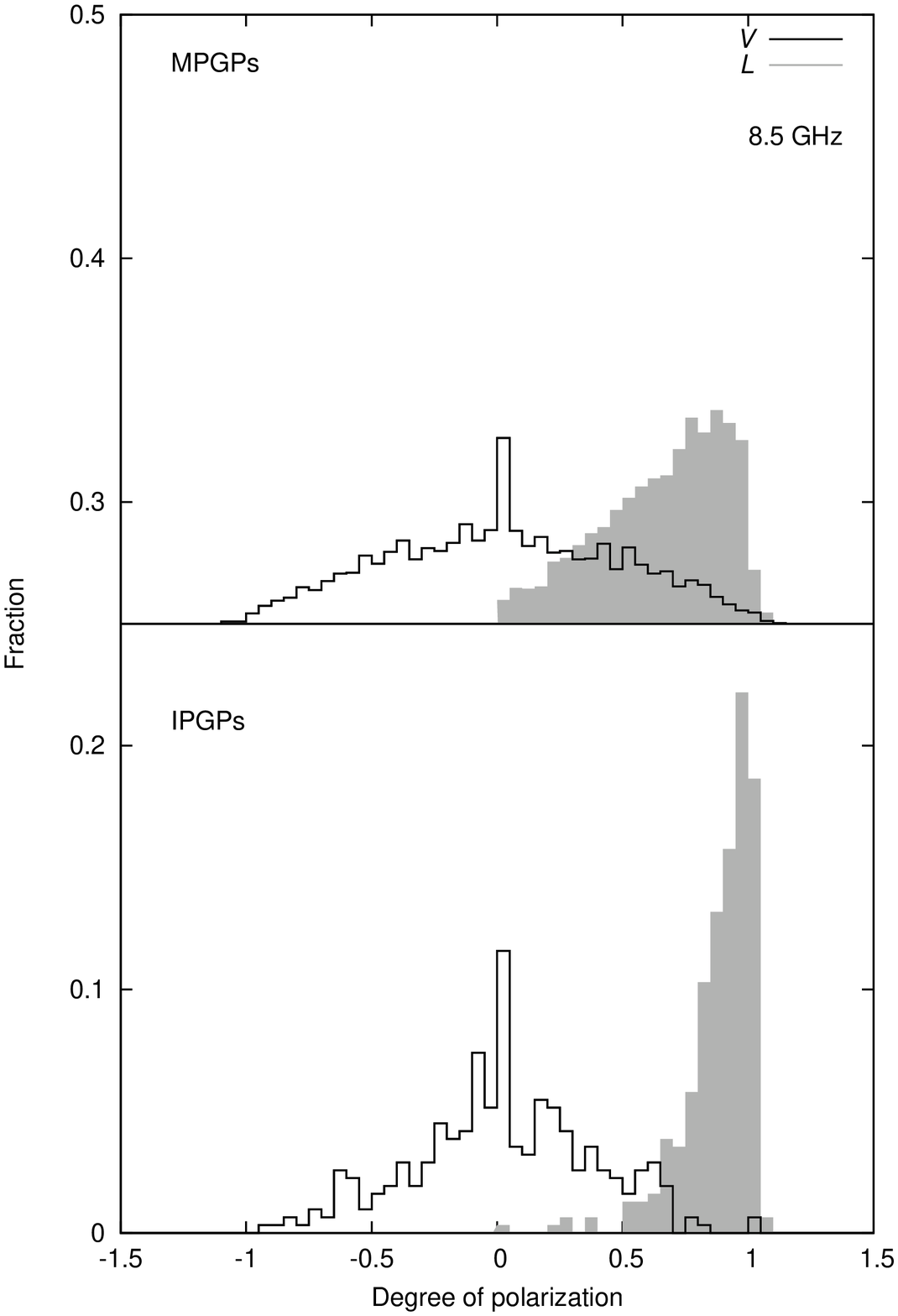}
\includegraphics[width=6.5cm,height=6cm]{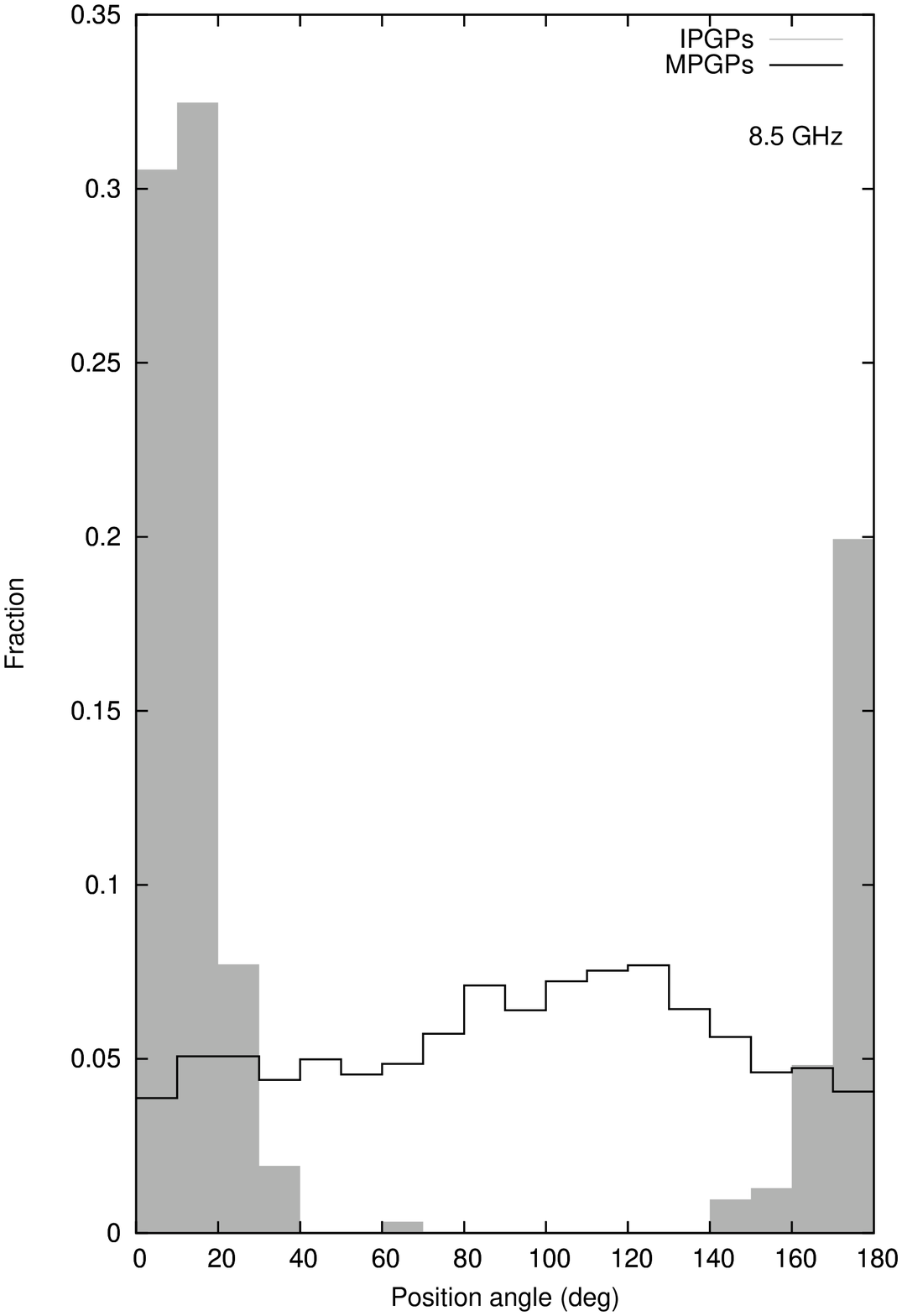}
\centering
\includegraphics[width=6.5cm,height=6cm]{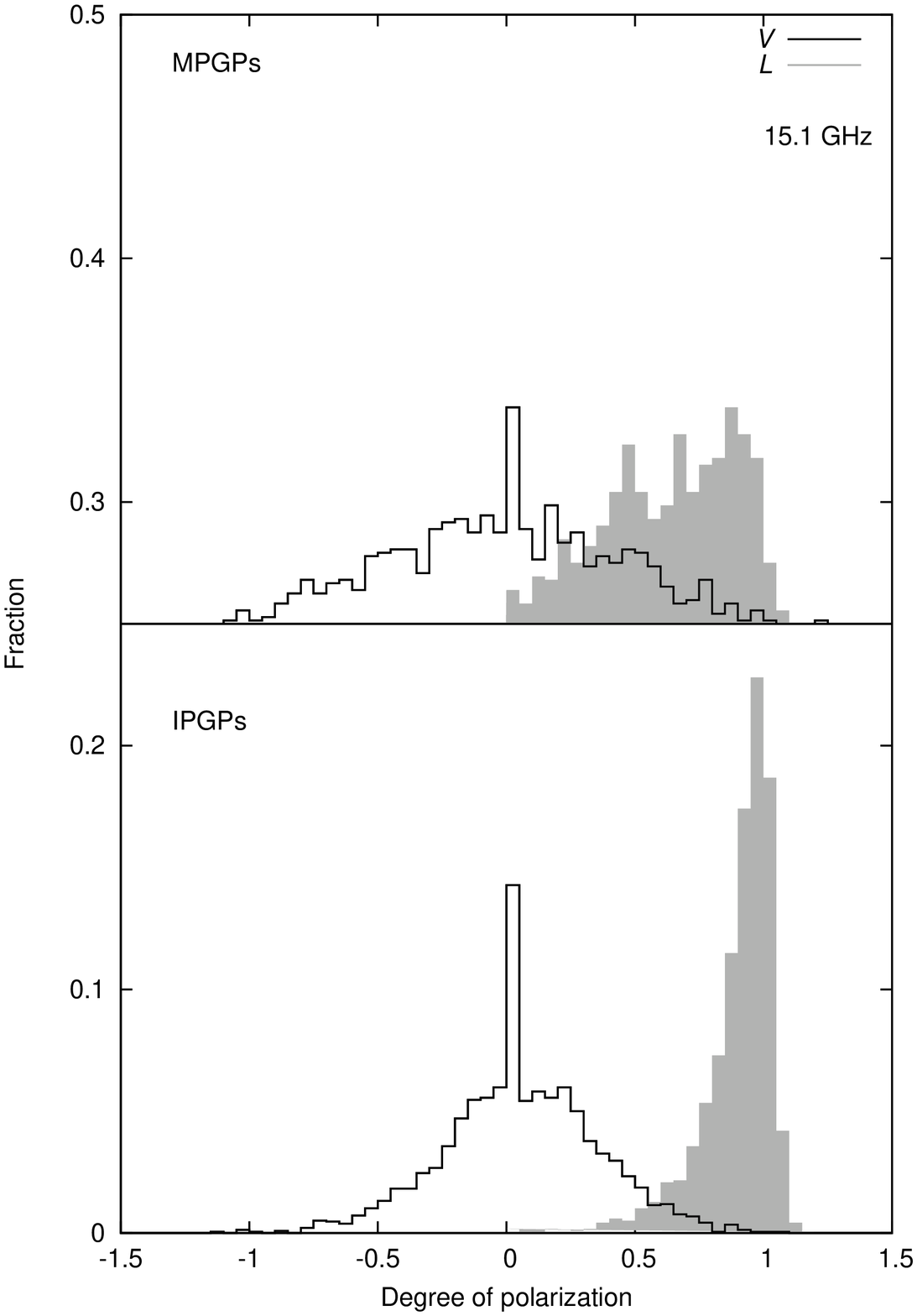}
\includegraphics[width=6.5cm,height=6cm]{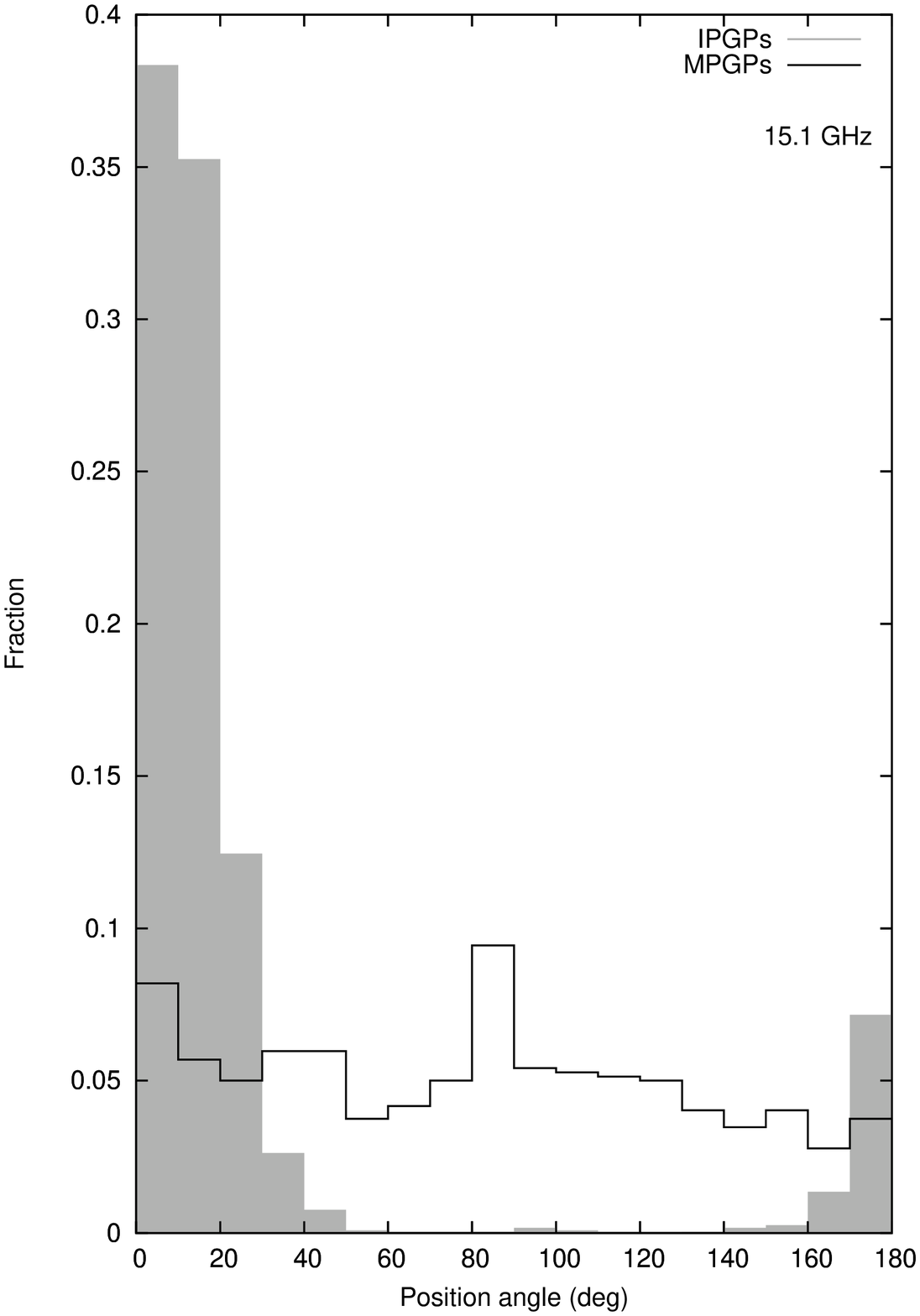}
\caption{Histograms of degree of polarization (left panels) and position angle of linear
polarization (right panels) for individual isolated strong spikes of radio
emission constituting every GP. Top panels correspond to GPs detected at 8.5~GHz,
and bottom panels~-- to GPs at 15.1~GHz. For histograms of degree of polarization,
open histograms represent the circular polarization, and filled histograms~-- linear
polarization.
}
\label{fig:pol_stat}
\end{figure*}

Some statistical properties of GP polarization are presented in Fig.~\ref{fig:pol_stat}.
Linear and circular polarizations, and position angle were calculated 
at the point of maximum total intensity $I$
for every spike above a given threshold in $I$
($10\sigma$ in signal smoothed by 5 samples).
There are 3250, 311, 720, and 2360
measurements for 
MPGPs and IPGPs at 8.5~GHz, and MPGPs and IPGPs at 15.1~GHz, respectively. 

One can see that the polarization properties of spikes which constitute
GP radio emission are nearly identical at both 8.5 and 15.1~GHz. MPGPs
demonstrate a rather broad distribution over the degree of circular and linear
polarization, and the PA of linear polarization shows a uniform distribution.
The distribution of the degree of polarization of IPGP spikes is narrower
with a greater portion of events having a high degree of linear polarization
and PA is concentrated in a restricted range of values between $-10\degr$
and $+20\degr$. It seems that there is also a distinct population of short
spikes with pure linear polarization, which produce a narrow maximum at $V=0$
in the distributions over the degree of circular polarization. 
The observed degree of linear and circular polarizations is quite different
from those observed for the GPs from the millisecond pulsar B1937+21 \citep{kondratiev2007},
where the majority of GPs were either highly circularly or linearly polarized, but the
degree of linear polarization did not exceed 0.6.

\subsection{Spectra of single pulses}
\begin{figure}[htb]
\centering
\includegraphics[scale=0.28,trim=0 0 150 20]{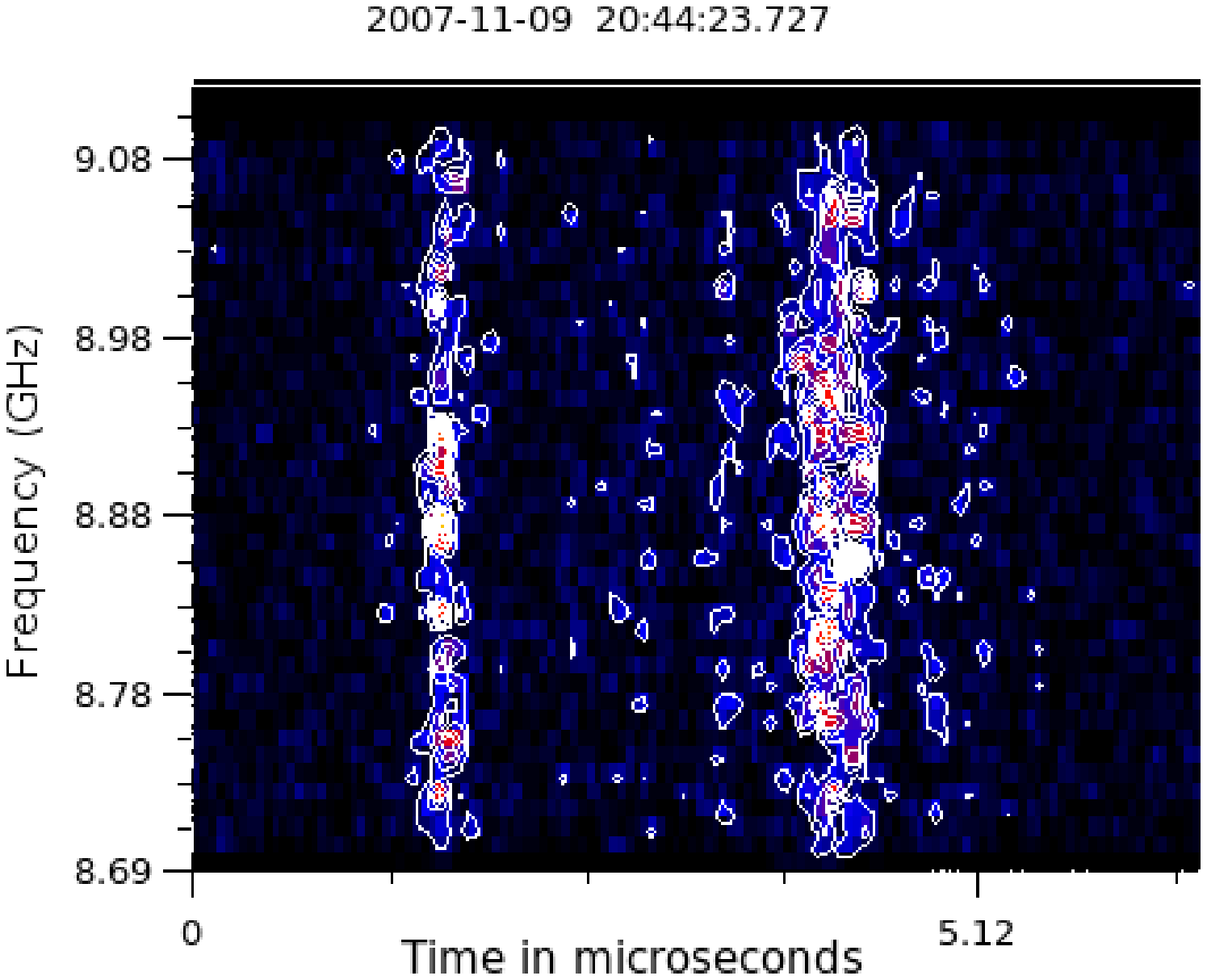}\includegraphics[scale=0.28,trim=0 0 150 20]{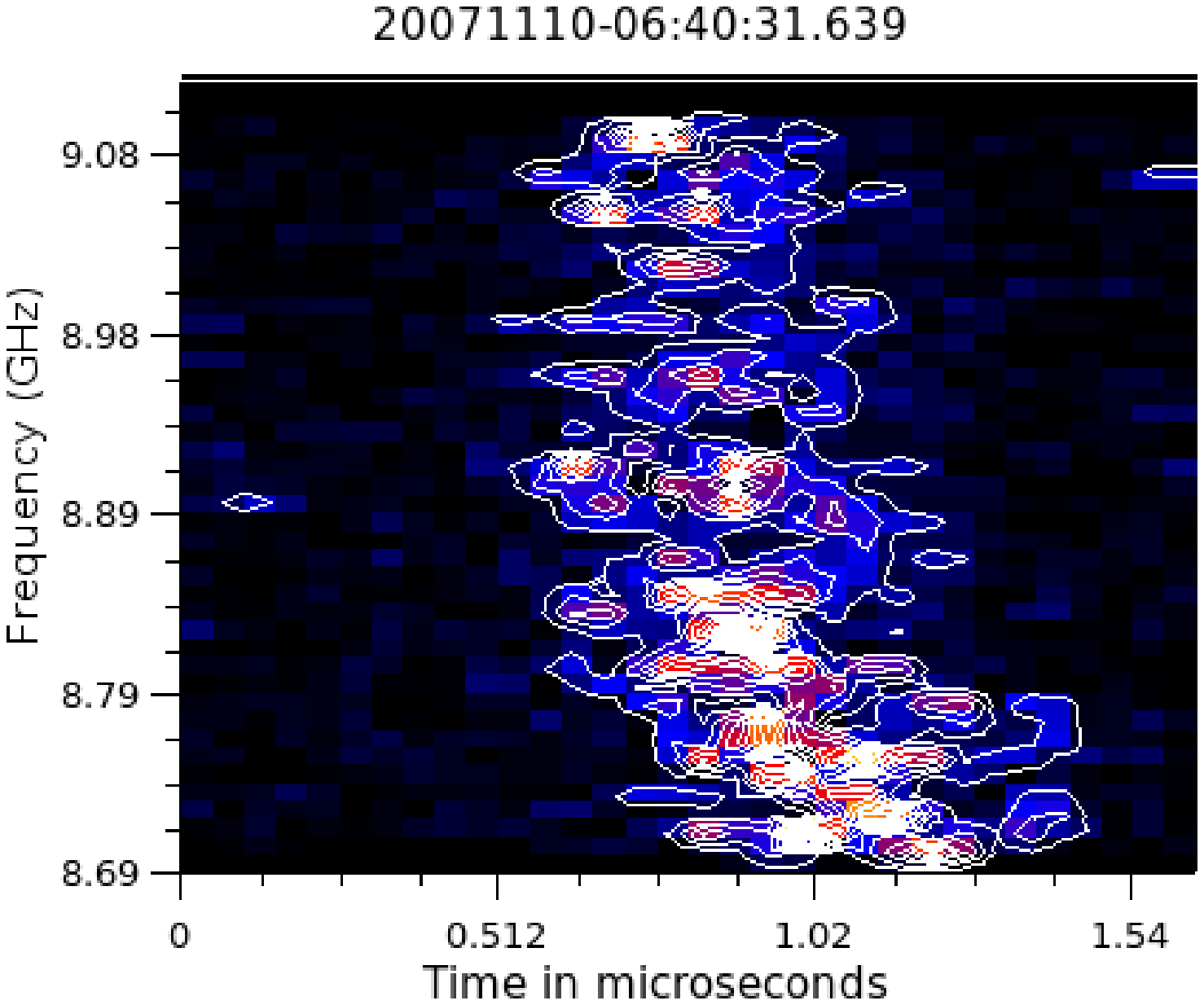}
\includegraphics[scale=0.28,trim=0 0 150 10]{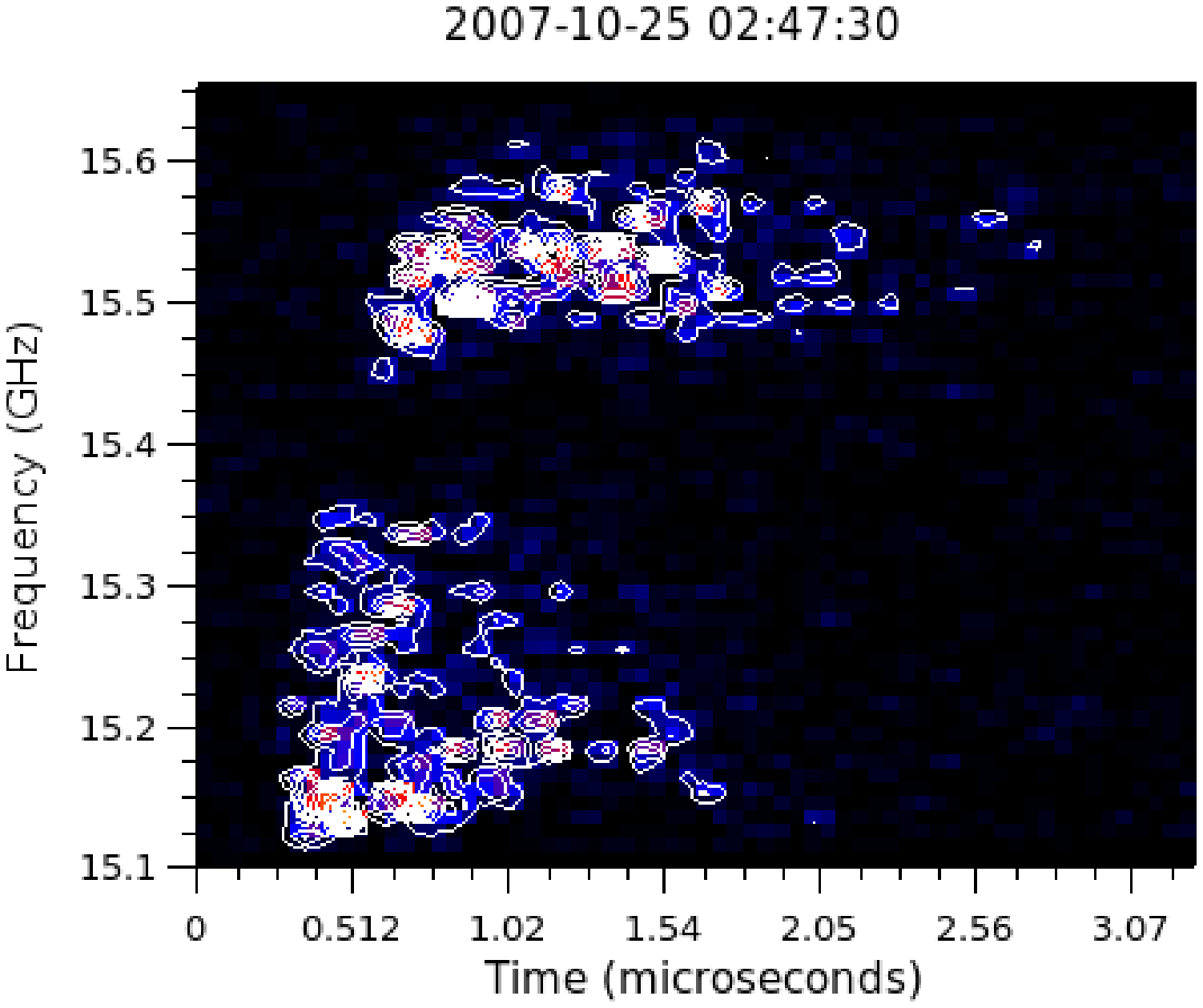}\includegraphics[scale=0.28,trim=0 0 150 10]{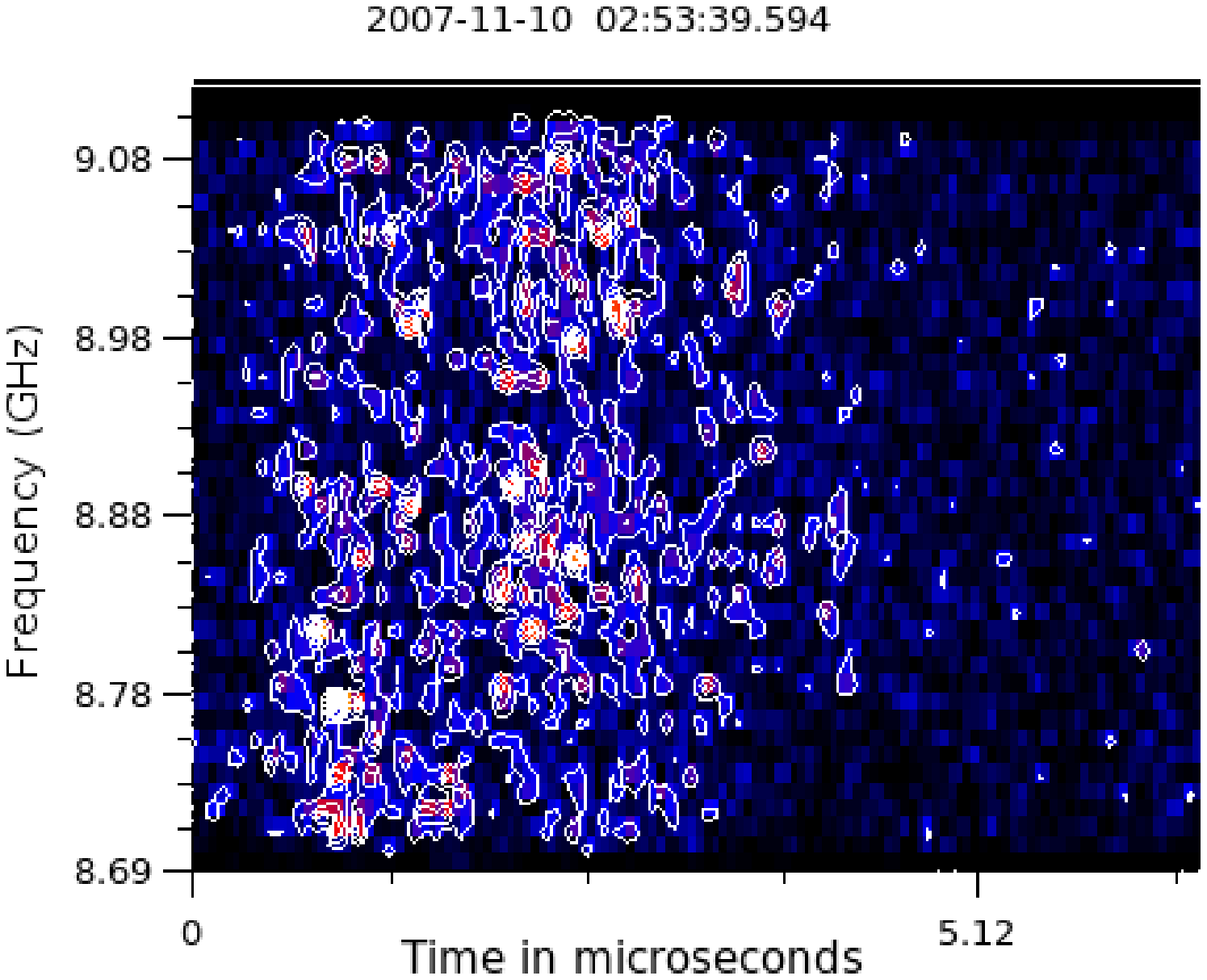}
\caption{Dynamic spectra of selected GPs. Left plots show the typical examples 
of dynamic spectra for MPGPs (top) and IPGPs (bottom). Right plots show the peculiar
GPs detected at 8.5~GHz at the longitude of MP (top, GP~639) and IP (bottom, GP~594).
}
\label{fig:spectra}
\end{figure}

To obtain dynamic spectra of the GPs we calculated the
detected  signal in narrow bands consisting
of 8192 samples corresponding to  9.765~MHz channel bandwidth
 after applying phase corrections
to the total spectrum. There are 44 and 55 such narrow bands in the dynamic
spectra at 8.5 and 15.1~GHz, correspondingly, 
with the 
sampling time 
being equal to 51.2~ns.
Examples of dynamic spectra for selected GPs are shown in Fig.~\ref{fig:spectra}.

\citet{hankins2007} found striking differences in the radio
spectra of both MPGPs and IPGPs. IPGPs have
emission bands proportionally spaced in
frequency with $\Delta\nu/\nu\approx0.06$, while MPGPs 
are characterized by broad-band spectra which smoothly fill 
the entire observing bandwidth.
Our observations confirm the presence of emission bands in IPGPs 
radio spectra at both 8.5 and 15.1~GHz (see Fig.~\ref{fig:spectra}, bottom-left). 
The total bandwidth we used (about 0.5~GHz)
does not allow us to measure the spacing between emission bands, since
the expected values are 0.5 and 0.9~GHz at 8.5 and 15.1~GHz, respectively.
The two emission regions seen in the bottom-left panel of Fig.~\ref{fig:spectra}
do not represent separate emission bands but constitute the internal structure
of an emission band typical for IPGPs, as was demonstrated by \citet{hankins2007}. 

\begin{figure}[hbt]
\centering
\includegraphics[scale=0.28,trim=0 0 150 20]{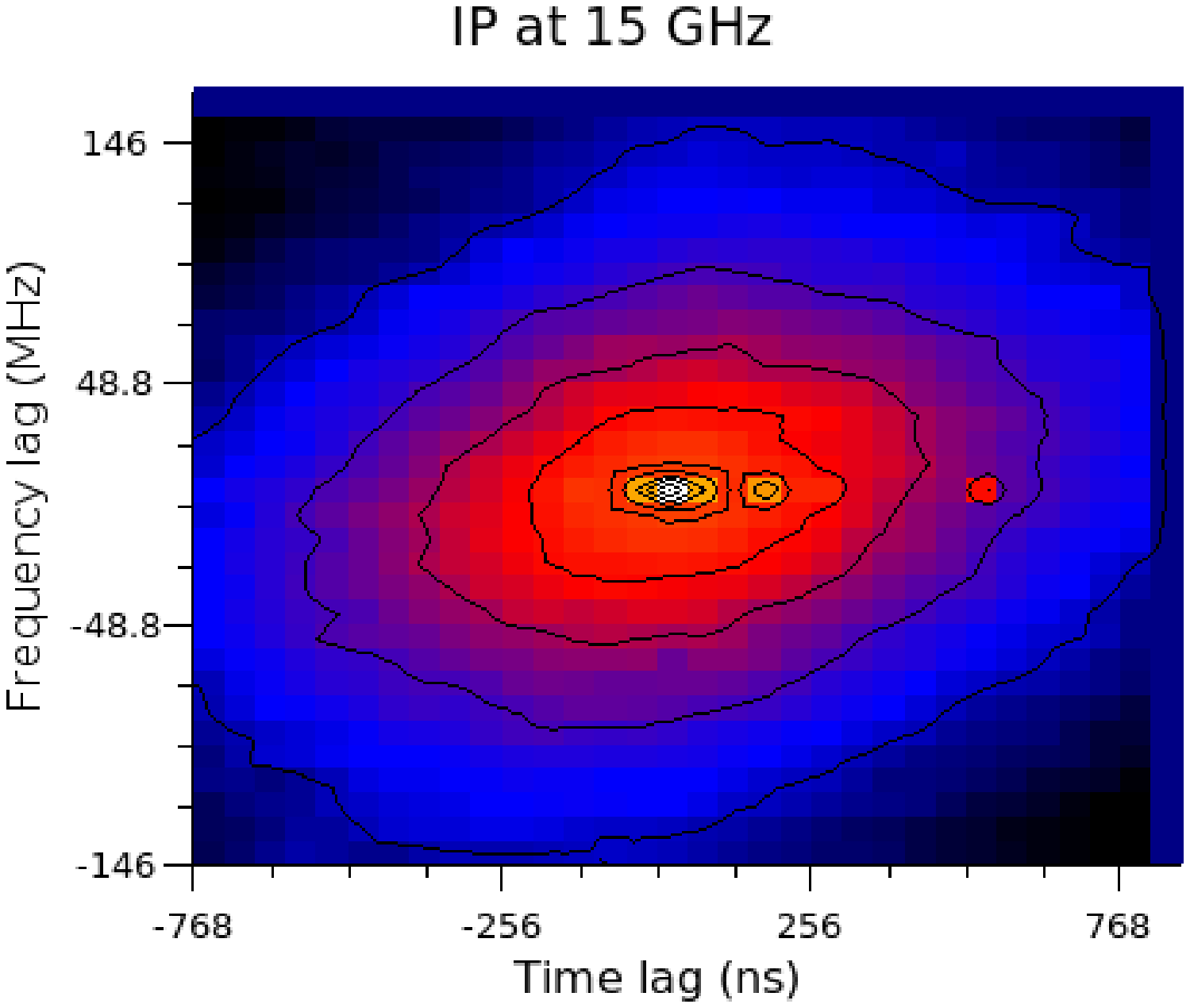}\includegraphics[scale=0.28,trim=0 0 150 20]{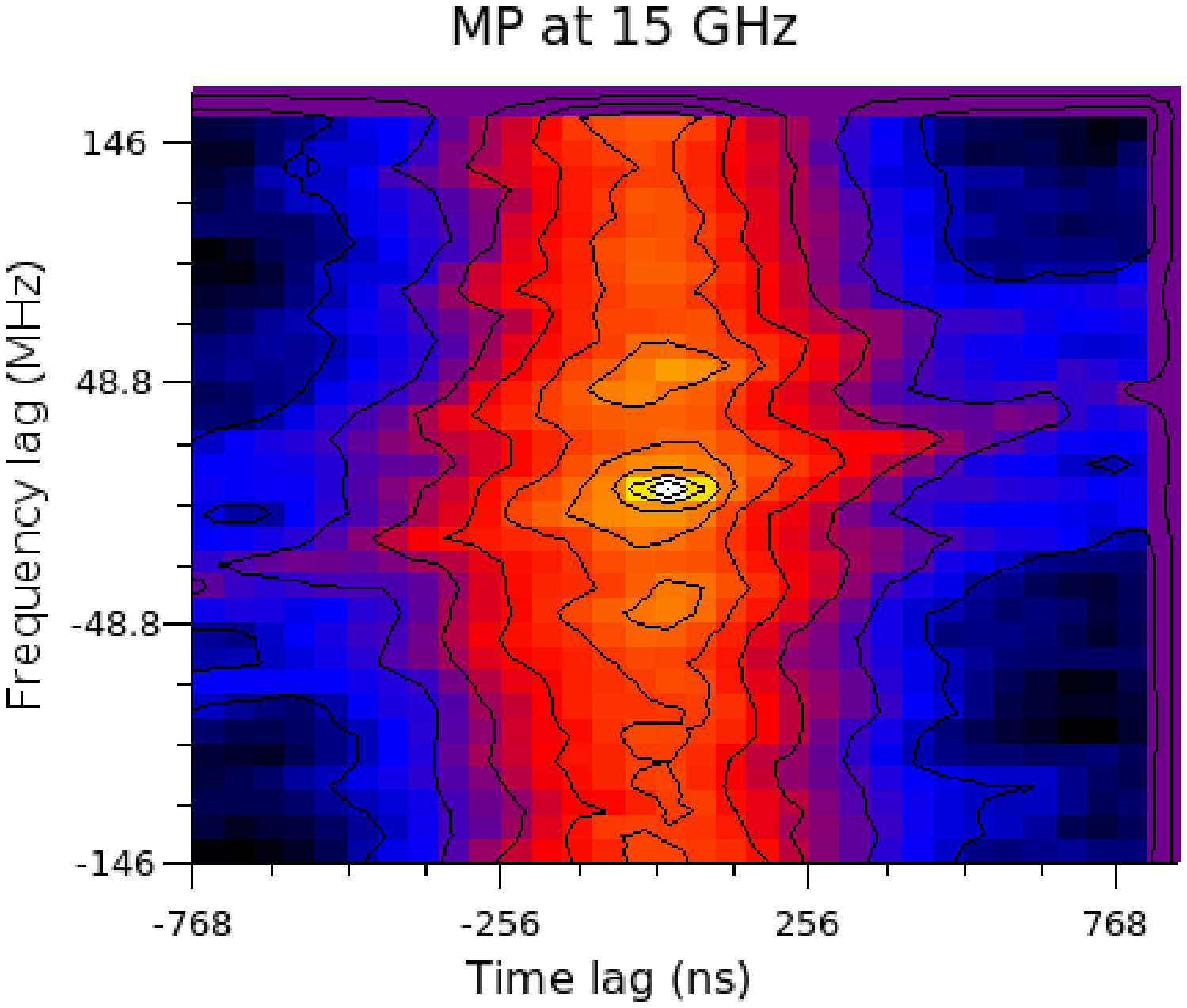}
\includegraphics[scale=0.28,trim=0 0 150 10]{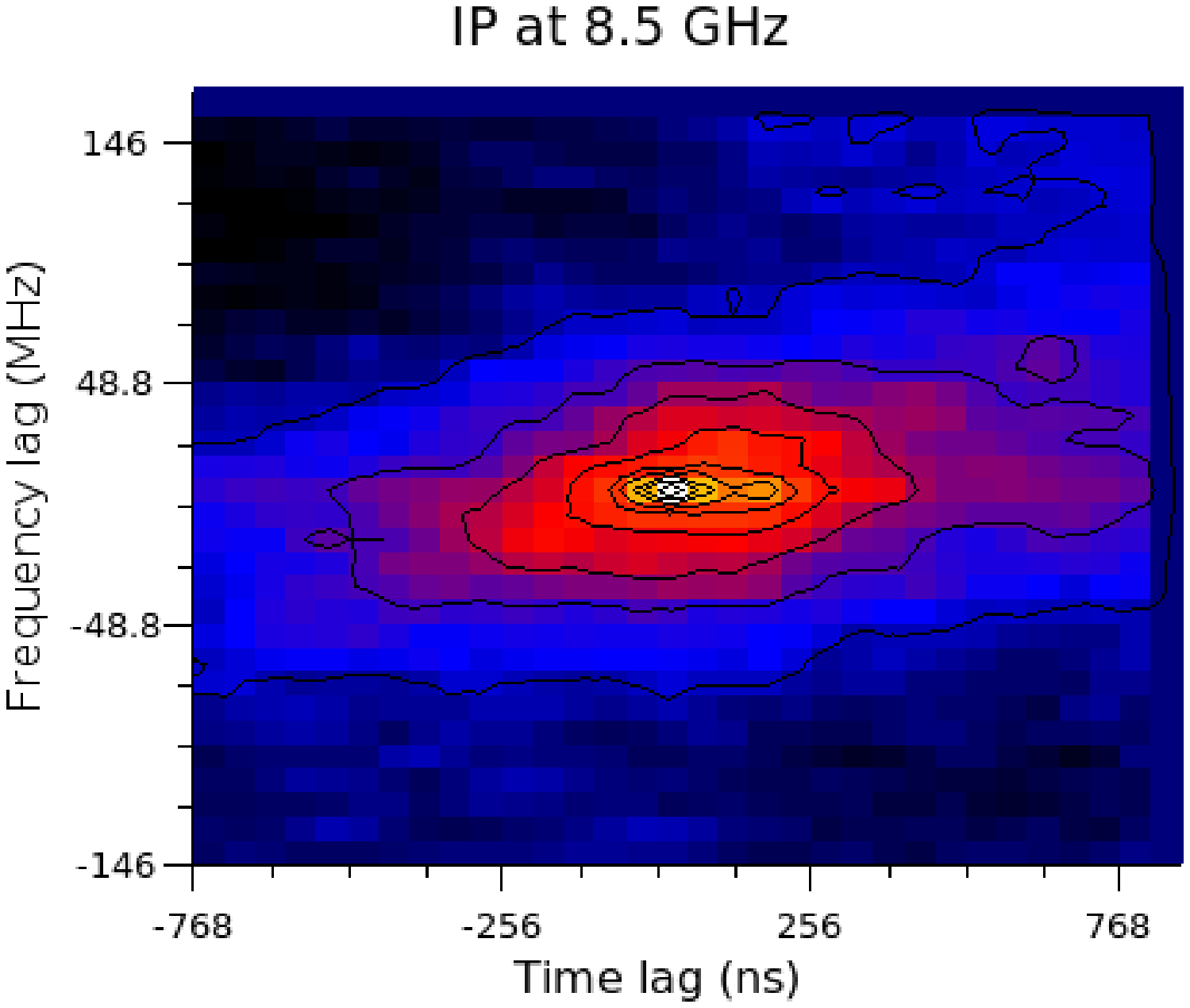}\includegraphics[scale=0.28,trim=0 0 150 10]{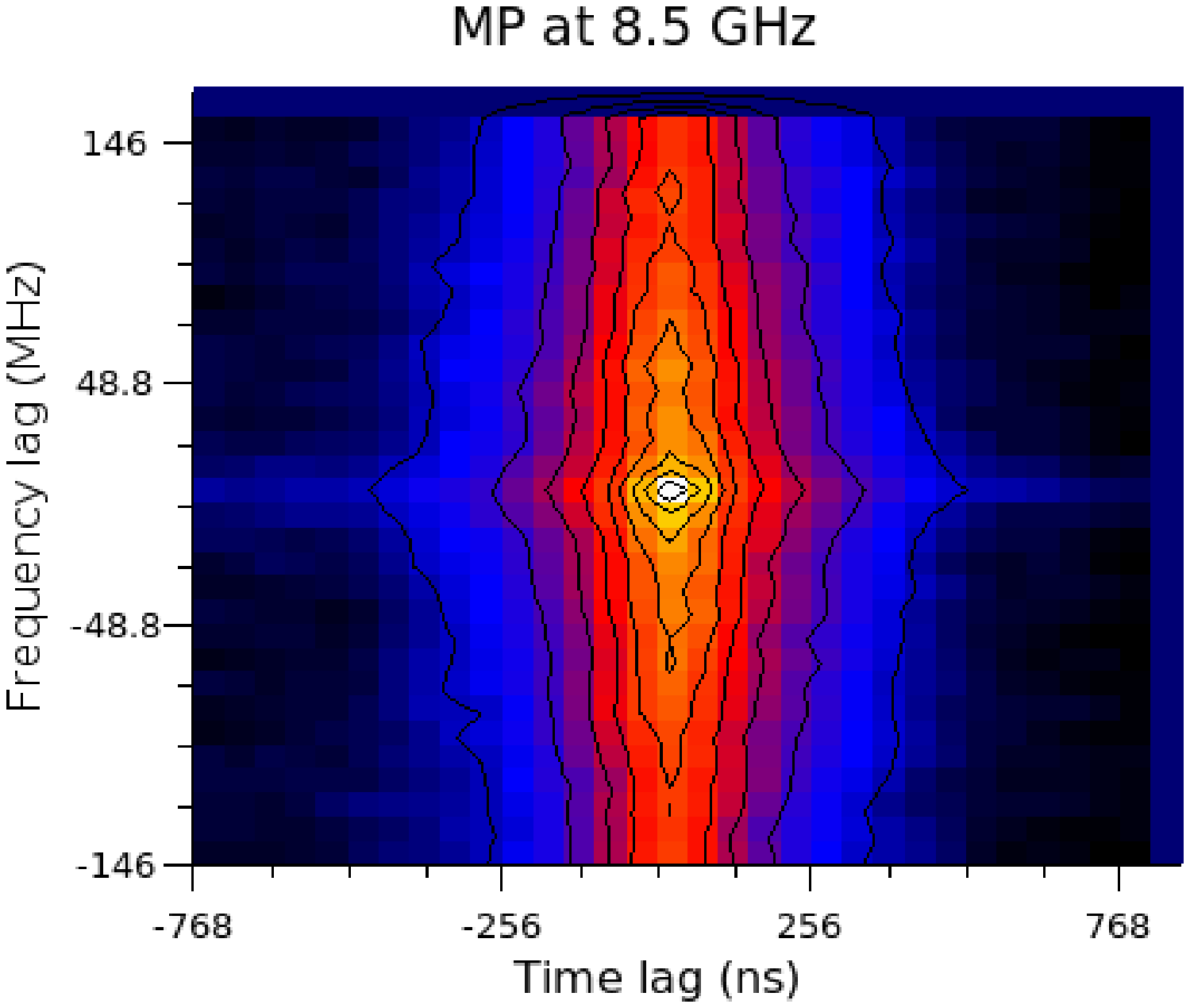}
\caption{Average 2-D CCFs between dynamic
spectra for RHC and LHC polarization channels at the frequency of 15.1~GHz (top) and 8.5~GHz (bottom)
for both IPGPs (left) and MPGPs (right). Contour levels are equidistant in steps of 10\% of the maxima given in fig. 13. 
}
\label{fig:two-dimCCF}
\end{figure}

\begin{figure}[ht]
\centering
\includegraphics[width=6cm,height=6cm]{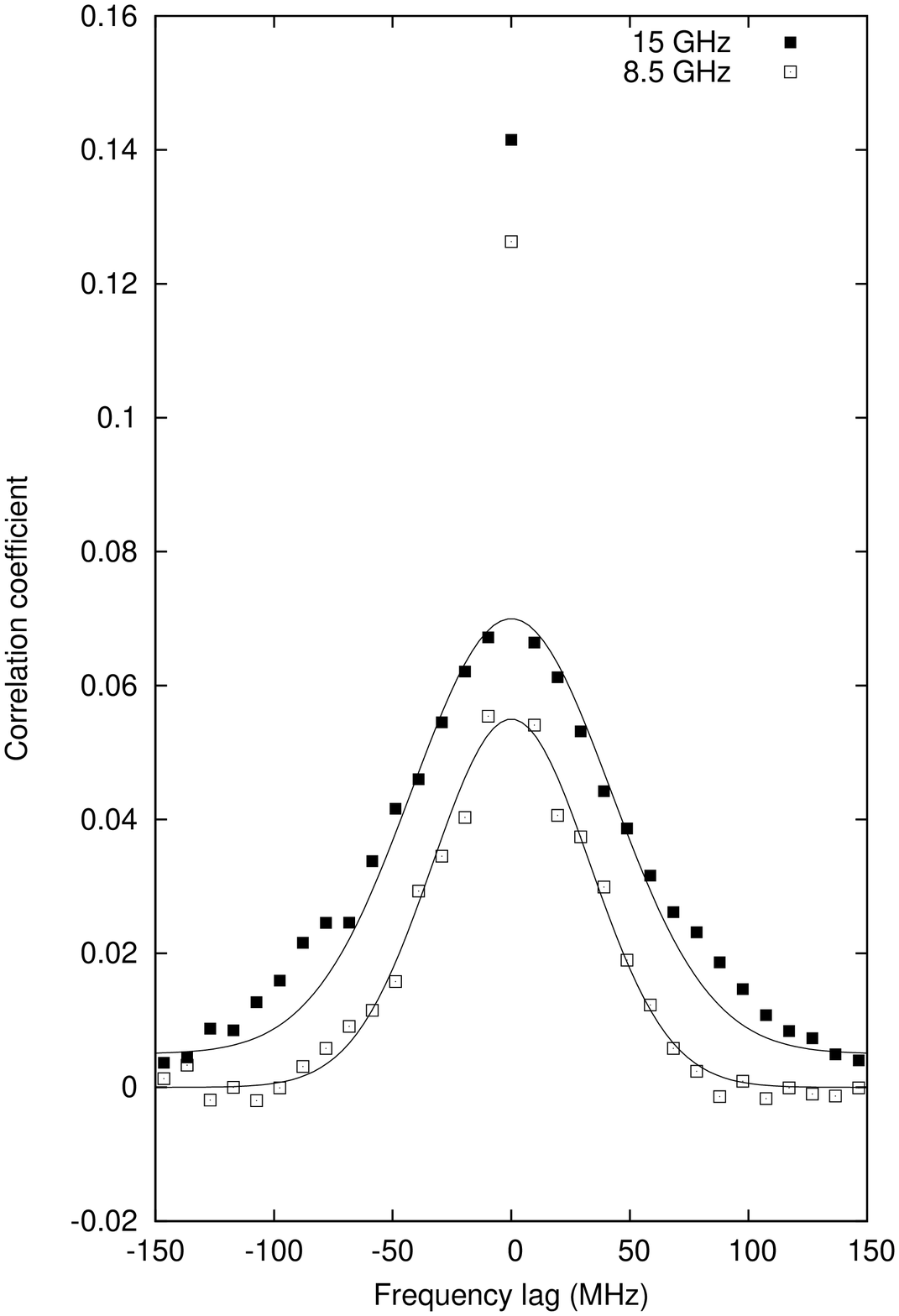}
\includegraphics[width=6cm,height=6cm]{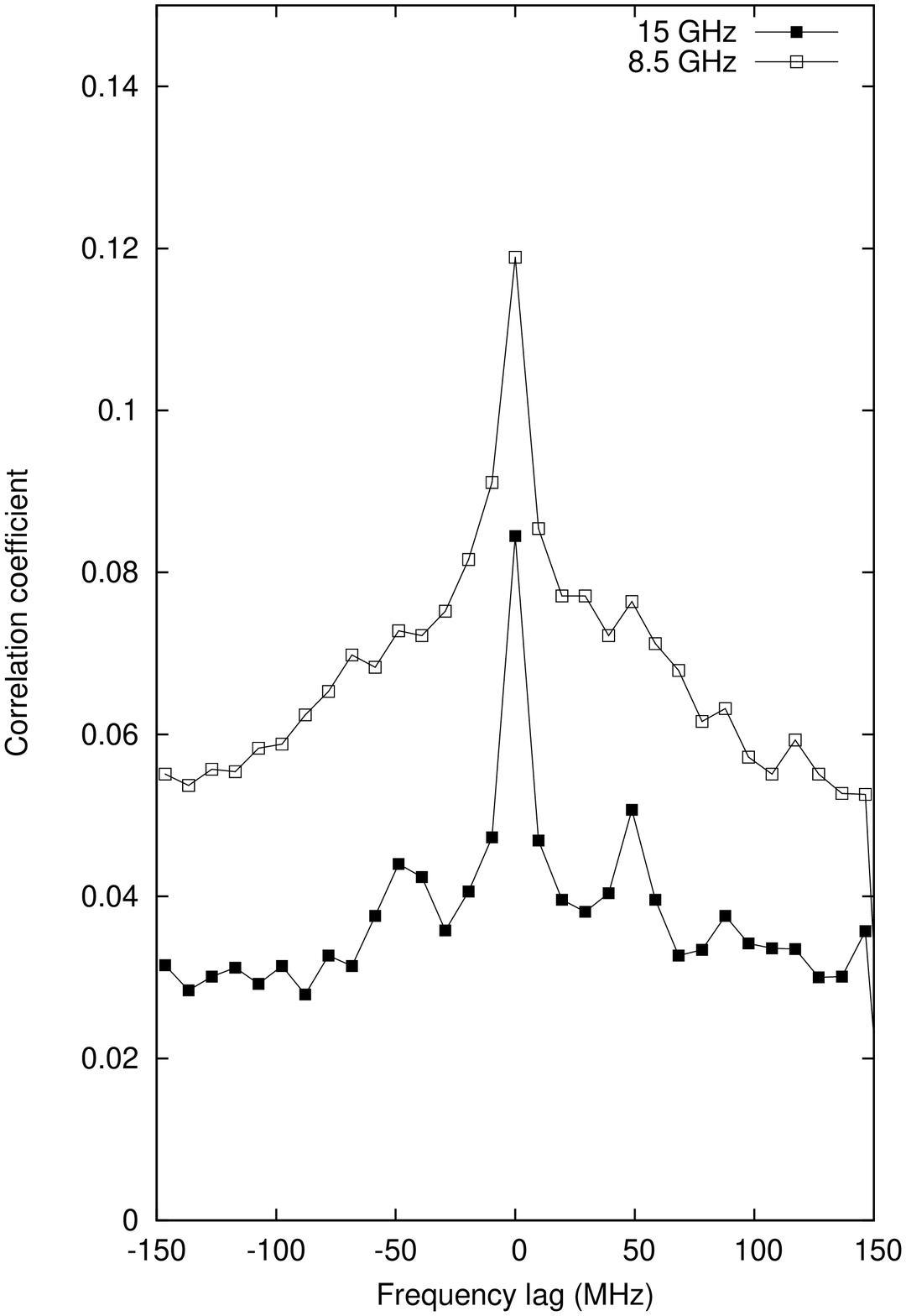}
\caption{Frequency cuts of 2-D CCFs from the Fig.~\ref{fig:two-dimCCF} between
dynamic spectra at the frequency of 8.5~GHz (open squares) and 15.1~GHz (filled squares) 
for IPGPs (top) and MPGPs (bottom).
}
\label{fig:freq_sec}
\end{figure}

We attempted to estimate the characteristic bandwidth
of such structures by analyzing the 2-D CCF between dynamic spectra
for RHC and LHC polarization channels. CCFs instead of ACFs were used in order
to reduce the contribution from the receiver noise at the ACF origin. 
Because of the large degree of linear
polarization such CCFs have a good signal-to-noise ratio, particularly for IPGPs.
Although the receiver noise was suppressed, the effect of correlated
noise from the strong background linear polarization remains.
The resulting  average CCFs are displayed in Fig.~\ref{fig:two-dimCCF}.
and show clear differences between IPGP and MPGP CCFs.
 There are broad components in the CCFs for the IPGP for which  Gaussians
with half-widths of
40 and 60~MHz can be fitted for 8.5 and 15.1~GHz respectively. The observed increase
in bandwidth is roughly proportional to frequency similar to the
effect seen in the separation between emission bands \citep{hankins2007}.
{
   Narrow repetitive features in the frequency direction 
with a period of about 40 MHz are evident in the CCFs of MPGPs 
at 8.5 and 15.1~GHz respectively. The pattern is stronger for MPGPs at 15.1 GHz.
In Fig.~\ref{fig:freq_sec} we  present the cuts of averaged CCFs for MPGPs
and IPGPs along the frequency axis at zero time lag which shows the feature quite clearly 
and enables the determination of their separation. 
The features are too strong to be of a background or instrumental 
origin  and if they were, they should also be visible in the background of the IPGP and be 
independent of the time lag which they are not, as seen in Fig.~\ref{fig:two-dimCCF} 
for 15.1 GHz MPGP. Closer inspection of the MPGP dynamic spectra (top of Fig.~\ref{fig:spectra}) 
reveals that there are also strong repetitive features in frequency when GP 
intensities are low. This is not the case for the IPGPs.  
}
Broad features cannot be distinguished as they
are comparable to the total receiver bandwidths (185 and 235~MHz
at 8.5 and 15.1~GHz, respectively).

\subsection{Peculiar giant pulses}
\label{sub:pec}
While in the previous sections we emphasized the differences in properties
of GPs belonging to MP and IP longitudes, a few examples 
of peculiar giant pulses were seen which violate general rules. 
One example is shown in Fig.~\ref{fig:smoothed_pol}, middle. 
For future reference we designate
this pulse as GP~639 using the fractional part of seconds in its arrival time as a label.
Seen at the MPGP longitude at 8.5~GHz, GP~639 has a smooth shape 
and high degree of linear and circular polarization otherwise typical of IPGPs.
Moreover, the degrees of linear and circular polarization are astonishingly similar in value
over the whole profile. These parameters are also similar for the average
profile of individual spikes constituting this GP (see Fig.~\ref{fig:spikes}, middle).
It is unlikely that such combination of a polarization properties was produced
by chance. GP~639 has also a distinct appearance in the dynamic spectrum 
with an unusual curvature
at the low frequency part of the band (Fig.~\ref{fig:spectra}, upper-right). 
Such a curvature in dynamic spectra was found by 
\citet{hankins2007} for IPGPs. 

Another example of a peculiar GP is presented by 
GP~594 detected at 8.5~GHz at the IP longitude (Fig.~\ref{fig:spectra}, bottom-right).
This pulse does not manifest any 
emission bands in its dynamic spectrum in contrast to the majority
of IPGPs (see bottom-left and bottom-right panels of Fig.~\ref{fig:spectra} for comparison).
It is interesting to note that the arrival longitude of GP~594 corresponds to that of IPs seen 
in the frequency range below 4~GHz. 
Thus, it appears that this pulse did not  undergo any longitude
shift and modulation of its radio spectrum.

\section{Discussion}
\label{disc}
In our study we have found a very sharp difference in properties of MPGPs and IPGPs
which are summarized below in Sect.~\ref{summary}. During our observations at both 8.5 and 
15.1~GHz no GPs at the longitudes of HFCs
were found. However, \citet{jessner2005} reported detecting GPs in all
phases of normal radio emission at the similar frequency of 8.35~GHz. 
In their experiment they
used the standard Effelsberg Pulsar Observation System (EPOS)
and a similar ultra-high time resolution oscilloscope (Le Croy LC584AL).
About 5\% of the GPs detected by  EPOS occurred at phases corresponding to the HFCs, the
GPs themselves were usually quite weak and only very few strong ones were detected.
The pulse phases of the oscilloscope data could not be determined with sufficient accuracy
to allow the identification of GPs occurring at HFC pulse phases. In the 2007 experiment
we detected 29~GPs at 8.5~GHz so that the predicted probability of detecting
at any GPs in the  HFC phase range is very low. As most GPs are below the trigger threshold
of the high resolution equipment, the absence of detected GPs in the HFC range at 8.5~GHz is not unexpected.
The problem is further aggravated at 15.1~GHz, where integrated profiles do not
show any HFC emission and GP emission probabilities of the Crab pulsar seem to follow the
integrated pulse profile.

\subsection{Rotation of polarization plane}
\label{sub:rot}

As was mentioned in Sect.~\ref{sub:pol}, distinct microbursts of MPGPs have 
position angles showing a classical smooth sweep that is very similar to that observed for 
integrated profiles of normal radio pulsars. This ``text-book'' S-sweep
is well-described by a geometric rotating vector model \citep[RVM,][]{rvm}.
Is it possible to explain apparent PA change in MPGPs with the RVM model? 
Formal fitting gives an improbably small value of the impact parameter of $0.1$--$20\arcsec$ 
from the magnetic axis. 
This corresponds to the linear distance of just 1--20~mm near the star's surface 
and 1--100~m at the light cylinder. 
This is a natural consequence of the fast PA variations, 
typically of $\sim 40$--$60\degr$ over about 500~ns, 
sometimes even faster. 
The strongest of the detected pulses observed at 8.5~GHz, 
consisting of two close very narrow spikes, 
shows a $20\degr$ smooth PA change over only 4~ns!

One has to note, however, that  for a rapidly rotating pulsar the magnetosphere 
differs greatly from that of a slowly rotating dipole anchored to a non-compact body. 
For a neutron star, any emission originating at heights of less than $1 R_\mathrm{NS}$ 
will be affected by the additional curvature of field lines caused by the Lens-Thirring effect.
A description of the field configuration near the surface of the star was  
given by \citet{muslimov1992}.
The simple RVM fails for another reason within the pulsar magnetosphere. It had been shown 
by \citet{dyks2003} 
as well as by \citet{wang2006}
that aberration and retardation distort the magnetic field lines and cause significant 
modifications to the PA curve for real pulsars. These effects need to be considered 
in any attempts to locate the GP within the magnetosphere. The fact that MPGPs and IPGPs 
have such different characteristics may point to propagation effects within the pulsar 
magnetosphere, where different ray paths will affect the signals differently, 
leading to additional scintillation and polarization effects.

However there are two simple heuristic ways to solve
the problem in the frame of RVM. First, let us suppose 
that the local magnetic field at the polar cap near the star's surface has a structure 
consisting of a number of narrow field tubes similar to that observed in solar spots. 
In this case the impact parameter is the distance from the center of the local flux tube. 
Individual microbursts are related to different tubes, the impact parameter is small but 
distinct for different microbursts. 
It is expected that the polar cap current will break up into individual filaments carrying 
currents  of order $1.7\times 10^4 \times \gamma^3$~A, forming small flux tubes as a consequence. 
The details of the radio emission from such flux tubes are the subject of a forthcoming paper.

The second possibility is that the PA sweep is related, not to the relatively slow rotation of the neutron star 
but to rapid movement of the GP emitter 
across the polar cap with the velocity of the order of the speed of light. 
The emitter then crosses a polar cap in different directions at different distances from the magnetic axis.
However, the exact nature of such an emitter is unknown at present.

\subsection{Emission mechanism}
\label{sub:em}
Though several hypotheses and models have been put forward to explain the GP phenomenon,
none of them can fully quantitatively address all the observed properties of GPs.
A number of attempts have been undertaken to implement plasma mechanisms for generation 
of GPs \citep{petrova2006, weatherall1998, lyutikov2007, istomin2004}. 
However, \citet{hankins2003} pointed out that the volume density of energy of GP emission 
is of the same order as the volume density of plasma energy and sometimes even exceeds it for 
the strongest GPs with peak flux densities of several million janskys,  
as recently reported by \citet{soglasnov2007} and \citet{hankins2007}. 
Moreover, a very high field strength of electromagnetic waves does
affect 
the interaction between the wave and particles \citep{soglasnov2007} significantly,
so that the mechanisms of linear plasma theory are no longer applicable.

It is very attractive to suppose that GPs are generated  directly by a polar cap discharge 
when cascade pair creation leads to a rapidly increasing volume charge and current.
Microbursts observed in MPGPs could be identified with sparks in the polar gap. 
IPGPs at high frequencies seem to be a rather stable phenomenon, and they can 
be described as ``well-mixed emission'' probably as a result of some scattering process
\citep[e.g. by inverse Compton scattering as suggested by][]{petrova2009}.
An alternative explanation of GP emission mechanism will be discussed in
detail by \citet{soglasnov2010}.
Here we sketch only a
very preliminary scenario of GP generation.

Originally GPs are created as a direct consequence of the polar cap discharge. 
A part of the energy of the GP is spent on particle acceleration. Because of the steep power-law 
radio spectrum, only the strongest GPs are detectable at high frequencies, while weak ones
(which mainly form an average profile) disappear completely.
This explains that in the case of MPGPs we see the faint but true original 
MP in their average profile at high frequencies, as well as strong but rare GPs 
at corresponding longitudes of the MP. 

The interpulse is weaker since it vanishes earlier at lower radio frequencies than the MP. 
What we see at higher frequencies is not the ``true'' IPGP, but
the emission of relativistic particles accelerated by the original IPGP.

{
The repetitive frequency pattern that we observed can be the result of a diffraction effect,
either from differing propagation paths in the pulsar magnetosphere or 
from the shape and structure of the coherent IP emission zone itself. 
The 40~MHz separation seen in the dynamic spectra of MPGPs would then correspond 
to a path difference of 7.5~m. The fact that the dynamic spectra
show regular but rather strong and narrow spikes of a width of less than 10 MHz
 over a weaker background can be interpreted as the emission from a succession of 
more than four very narrow discharge zones (unresolved width $< {c \over \rm 500~MHz} = \rm 0.6~m$) 
along the line of sight. In that case, one would also expect these features to be more clearly visible 
at higher frequencies, because at higher frequencies the phase of the radio emission from an individual 
narrow discharge zone will be more clearly defined, giving stronger interference patterns.
Additional support for the existence of nanoshot discharges can be obtained from high frequency observations 
of ordinary pulsars. The typical shape of their spectra may be interpreted
as being caused by a superposition of radiation from such nanoshot discharge regions \citep{Loehmer2008}.  } 

\section{Summary}
\label{summary}
A very striking difference in properties of MPGPs and IPGPs was revealed 
in our study.

\subsection{IPGPs}
\label{sub:ipl}
\begin{enumerate}
\item Waveform: IPGPs are always smooth in shape and typically asymmetric, with a
rather sharp leading edge with a rise time of $0.6\pm 0.2~\mu$s and gradual decay 
of about $2.5~\mu$s at the trailing edge.
\item Internal emission structure: IPGPs are filled with pure noise well described by 
the AMN model \citep{rickett1975}. 
\item Polarization: IPGPs exhibit a high degree of linear polarization with essentially 
a constant PA, restricted in the range of $\pm 5\degr$, similar at both 8.5 and 15.1~GHz.
\item Spectra: 
Radio spectra of IPGPs consist of emission bands
at 8.5 and 15.1~GHz as was first reported by \citet{hankins2007}. 
The half-width of the emission bands was found to be equal to
40 and 60~MHz for 8.5 and 15.1~GHz, respectively. 

\end {enumerate}

\subsection{MPGPs}
\label{sub:mpl}
\begin{enumerate}
\item Waveform: MPGPs demonstrate a large variety of shapes containing one or several 
microbursts of emission with a duration of less than a microsecond.
The bursts can  occur intermittently at random time intervals of several microseconds duration. 
The total time envelope of a given GP can extend over hundreds of microseconds. Microbursts can
often contain isolated or overlapping unresolved nanoshots of great intensity.
\item Internal emission structure: MPGPs are composed of distinct unresolved 
small spikes.
\item Polarization: MPGPs show a significant diversity of polarization parameters, with 
linearly and circularly polarized spikes being present in roughly equal proportions. 
The distribution of linearly polarized spikes (nanoshots) over PA looks uniform 
in the whole range of 0--$180\degr$. In the case of well-separated microbursts, 
PA demonstrates a rapid but smooth regular variation inside a microburst, 
very similar to that observed for integrated profiles of many pulsars. 
\item Spectra: { Dynamic spectra of MPGPs are broad-band, filling the entire observing 
bandwidth of 0.5~GHz. MPGPs show additional regular spiky frequency patterns with a separation of about 
40 MHz in their dynamic spectra. These also
show up in CCF between LHC and RHC polarization signals.}
\end {enumerate}

Such sharp contrasts between MPGPs and IPGPs are not observed at lower frequencies.
On the other hand it seems hardly probable that the emission mechanism of GPs is very 
different at high and low frequencies, 
hence the reason for the widely different GP characteristics at different pulse phases remains unknown.

\begin{acknowledgement}
This work is based on observations with the 100-m telescope of the
MPIfR (Max-Planck-Institut f\"ur Radioastronomie) at Effelsberg.
We are grateful for the technical support by the Effelsberg system group and operators
and the helpful comments by Ramesh Karuppusamy.
This work was partially supported by the Russian Foundation for Basic Research, 
through grant 07-02-00074. VIK is supported by a WV EPSCOR Research Challenge Grant.
YYK is supported by the Alexander von~Humboldt return fellowship.
\end{acknowledgement}

\bibliographystyle{aa}
\bibliography{v13.bib}

\end{document}